\documentclass[conference]{IEEEtran}
\IEEEoverridecommandlockouts

\usepackage{cite}
\usepackage{amsmath,amssymb,amsfonts}
\usepackage{algorithmic}
\usepackage{graphicx}
\usepackage{textcomp}
\usepackage{xcolor}
\usepackage[norelsize,ruled,vlined,linesnumbered]{algorithm2e}
\usepackage{algorithmic}
\usepackage{amsthm}
\usepackage{amsmath,amssymb,amsfonts}
\usepackage{booktabs}
\usepackage{url}
\usepackage{subfigure}
\usepackage[colorlinks=true,linkcolor=blue,citecolor=blue]{hyperref}
\usepackage[capitalize]{cleveref}

\newtheoremstyle{mystyle}
    {1.0mm}
    {1.0mm}
    {\it}
    {0.0mm}
    {\scshape}
    {.}
    { }
    {}
\theoremstyle{mystyle}
\newtheorem{definition}{Definition}
\newtheorem{example}{Example}
\newtheorem{lemma}{Lemma}
\newtheorem{theorem}{Theorem}

\newcommand{\vs}{\vspace{1.0mm}}
\newcommand{\wsq}{\hspace{\fill}$\square$}
\SetKwComment{Comment}{$\triangleright$\ }{}

\def\BibTeX{{\rm B\kern-.05em{\sc i\kern-.025em b}\kern-.08em
    T\kern-.1667em\lower.7ex\hbox{E}\kern-.125emX}}
\begin{document}

\title{Random Sampling over Spatial Range Joins}

\author{
\IEEEauthorblockN{Daichi Amagata}
\IEEEauthorblockA{\textit{Graduate School of Information Science and Technology} \\
\textit{The University of Osaka}\\
Japan \\
amagata.daichi@ist.osaka-u.ac.jp}
}

\maketitle

\begin{abstract}
Spatial range joins have many applications, including geographic information systems, location-based social networking services, neuroscience, and visualization.
However, joins incur not only expensive computational costs but also too large result sets.
A practical and reasonable approach to alleviating these issues is to return random samples of the join results.
Although this is promising and sufficient for many applications involving spatial range joins, efficiently computing random samples is not trivial.
This is because we must obtain random join samples without running spatial range joins.

We address this challenging problem for the first time and aim at designing a time- and space-efficient algorithm.
First, we design two baseline algorithms that employ existing techniques for random sampling and show that they are not efficient.
Then, we propose a new data structure that can deal with our problem in $\tilde{O}(n + m + t)$ expected time and $O(n+m)$ space, where $n$ and $m$ are the sizes of two point sets and $t$ is the required number of samples.
We conduct extensive experiments using four real spatial datasets, and the results demonstrate that our algorithm is significantly faster than the baselines in most tests.
\end{abstract}


\section{Introduction}  \label{sec:introduction}
Location-aware services are ubiquitous, and they obtain many geospatial points (i.e., points with x-y coordinates).
Analyzing spatial points is important to improve service quality and user satisfaction \cite{mamoulis2022spatial}.
As analytical operations, spatial joins are employed in many applications, such as geographic information systems \cite{georgiadis2023raster,nishio2020lamps,tsuruoka2020distributed,hori2023learned,amagata2021feat}, location-based social networking services \cite{yang2020lbsn2vec++,nishio2017geo}, entity linking \cite{shahvarani2021distributed}, neuroscience \cite{pavlovic2016transformers,amagata2019identifying}, clustering \cite{amagata2021fast,amagata2022scalable,amagata2023efficient}, outlier detection \cite{amagata2022fast,amagata2021fast_}, and visualization \cite{yu2021geosparkviz,amagata2022learned}.
Given an orthogonal rectangle (or window) centered at a point $r$, denoted by $w(r)$, we use ``$w(r) \cap s$'' if a point $s$ exists in $w(r)$.
Given two point sets $R$ and $S$, a spatial range join $R \Join S$ computes a set $J = \{(r,s) \,|\, r \in R, s \in S, w(r) \cap s\}$, where each $w(r)$ has the same window size \cite{vsidlauskas2014spatial,sowell2013experimental}.
($R$ and $S$ are symmetric and interchangeable.)
Because $R$ and $S$ are usually large, spatial range joins incur not only expensive computational costs but also substantial result sets.

\vs
\noindent
\textbf{Motivation.}
Naturally, the above applications want to avoid such time-consuming operations and overwhelming result sets.
A practical and reasonable approach to avoiding these concerns is to do random sampling from the join results \cite{haas1999ripple,acharya1999join,kim2023guaranteeing,deng2023join,chaudhuri1999random,chen2020random,huang2019joins}, because random samples of $J$ are usually sufficient to illustrate the distribution of $J$.
For example, when analyzing join results (e.g., statistics) for knowledge discovery, (kernel) density visualization, and spatial aggregation, random samples are sufficient to obtain (approximate yet) highly accurate results \cite{zhao2018random,shanghooshabad2021pgmjoins,li2016wander,li2019wander,li2017wander,zacharatou2017gpu,wang2015spatial,shekelyan2023streaming}.
In addition, the recent trend of AI/ML for databases also motivates join sampling.
For example, learned models for cardinality estimation \cite{yang2020neurocard,kim2022learned,heimel2015self} and query optimization \cite{vu2021learned,vu2024learning} on spatial databases are trained on random samples of join results.
It is important to notice that, to enable effective analysis (e.g., training learned models), random samples should be \textit{uniform} and \textit{independent}.

Although the problem of random sampling over spatial range joins has these important applications, this problem has not been tackled so far.
(\cref{sec:preliminary} formally defines this problem.)
This paper addresses this problem and aims to design a theoretically and empirically time- and space-efficient algorithm that yields uniform and independent random samples of spatial range joins.

\vs
\noindent
\textbf{Challenge.}
Solving this problem \textit{efficiently} is not trivial, and the time complexity of algorithms for the problem should be less than $O(nm)$, where $n = |R|$ and $m = |S|$.
Notice that $|J|$ can be as large as $O(nm)$, and thus $O(nm)$ time is equivalent to computing the full join result.
Therefore, the most straightforward algorithm, which runs a state-of-the-art spatial range join algorithm and then does random sampling from $J$, is infeasible.
Another intuitive approach is to prepare random samples of $R$ and $S$ (denoted by sample$(R)$ and sample$(S)$) and then consider sample$(R) \Join$ sample$(S)$.
However, as shown in \cite{chaudhuri1999random}, sample$(R) \Join$ sample$(S) \neq$ sample$(R \Join S)$, meaning that this approach cannot satisfy the uniform and independent requirements.
Uniform and independent join sampling has been studied mainly in the context of relational databases \cite{deng2023join,chen2020random,zhao2018random,kim2023guaranteeing,dai2024reservoir,wang2024join}, and they consider only equi-joins, which are totally different from spatial range joins.

To overcome the challenge of designing an algorithm that provides uniform and independent random samples without running a spatial range join, we first show that two existing techniques are available.
As a baseline, we consider KDS \cite{xie2021spatial}, which was originally designed for spatial independent range sampling (i.e., random sampling from an orthogonal range search result).
We specifically show that combining KDS with range counting can provide $t$ uniform and independent join samples in $O((n+t)\sqrt{m})$ time.
In this algorithm, range counting incurs $O(n\sqrt{m})$ time, which can be huge for large $n$ and $m$.
The idea of rejection sampling in \cite{zhao2018random} can alleviate this issue, and it shows that upper-bounds of the range counts still guarantee uniform and independent join samples.
A simple adaptation of grid mapping enables $O(1)$ time upper-bound computation for each point.
By fusing KDS, rejection sampling, and grid mapping, we obtain an $O(n+m+\frac{nm^{1.5}t}{|J|})$ expected time algorithm, which is used as another baseline.

Although the above two algorithms do not suffer from $O(nm)$ time, they still have several drawbacks.
First, as our experimental results demonstrate, KDS, which employs a $k$d-tree \cite{bentley1975multidimensional}, is inefficient.
This is because it requires $O(\sqrt{m})$ time for range counting and for sampling a point to satisfy uniform and independent join sampling.
Secondly, the approach based on rejection sampling works well only when upper-bounds have an approximation guarantee.
The simple grid adaptation does not guarantee any bound for approximate range counting, which may incur a low acceptance probability in rejection sampling, resulting in a worse performance than that of KDS.
Now we have a question:
\begin{quote}
\setlength{\leftskip}{-4.0mm}
\setlength{\rightskip}{-4.0mm}
Is there a data structure that (i) consumes only $O(m)$ space (to scale well to large datasets) and, for a given point, provides (ii) an upper-bound of range count with an approximation guarantee in $\tilde{O}(1)$ time and (iii) a random $(r,s)\in J$ in $\tilde{O}(1)$ time?    
\end{quote}
($\tilde{O}(\cdot)$ hides any polylog factors.)
Addressing this question is the main challenge of our problem, and it is not trivial to use existing range counting data structures (e.g., \cite{chan2016adaptive,rahul2017approximate,shekelyan2021approximating}) for solving the problem of sampling random pairs in $J$.
That is, data structures satisfying (i) to (iii) simultaneously are not known.

\vs
\noindent
\textbf{Contribution.}
This work provides a positive answer to the above challenging question by proposing a novel algorithm for the problem.
We specifically propose BBST (Bucket-based Binary Search Tree), a new data structure based on a binary search tree that (i) consumes only $O(m)$ space, (ii) runs an $\tilde{O}(1)$-approximate range counting in $\tilde{O}(1)$ time, and (iii) randomly samples $(r,s) \in J$ in $\tilde{O}(1)$ expected time.
In addition, we exploit the grid structure differently from the second baseline to avoid losing any approximation bounds.
As introduced later, the idea of BBST is to convert a 4-sided (window) range into a 2-sided one, which is available by using the grid structure.
Thanks to these techniques, we obtain an $\tilde{O}(n+m+t)$ expected time algorithm for obtaining $t$ uniform and independent samples of $J$.
\cref{tab:complexity} compares the time and space complexities of our algorithm and baseline ones.

To summarize, this paper makes the following contributions\footnote{This is accepted version, and the publication version is \cite{amagata2025random}.}.
\begin{itemize}
    \setlength{\leftskip}{-2.0mm}
    \item   We formulate the problem of random sampling over spatial range joins for the first time (\cref{sec:preliminary}).
    \item   We design two baseline algorithms for this problem (\cref{sec:baseline}).
    \item   We propose an $\tilde{O}(n+m+t)$ expected time algorithm that exploits a new data structure BBST (\cref{sec:proposal}).
    \item   We conduct extensive experiments on four real spatial datasets (\cref{sec:experiment}).
            The experimental results demonstrate that our algorithm is at least one order of magnitude faster than the baseline algorithms in most tests.
\end{itemize}

\begin{table}[!t]
    \centering
    \caption{Comparison of time and space complexities}
    \label{tab:complexity}
    \vspace{-2.0mm}
    \begin{tabular}{llc} \toprule
                                    & Time                                      & Space     \\ \midrule
        KDS \cite{xie2021spatial}   & $O((n+t)\sqrt{m})$                        & $O(n+m)$  \\
        KDS-rejection               & $O(n+m+\frac{nm^{1.5}t}{|J|})$ expected   & $O(n+m)$  \\
        BBST (ours)                 & $\tilde{O}(n+m+t)$ expected               & $O(n+m)$  \\ \bottomrule
    \end{tabular}
    \vspace{-2.0mm}
\end{table}

\section{Preliminary}   \label{sec:preliminary}
We use $R$ and $S$ to represent two spatial datasets, where $|R| = n$ and $|S| = m$.
As with existing spatial database works \cite{xie2021spatial,sowell2013experimental,doraiswamy2020gpu,zacharatou2017gpu,vsidlauskas2014spatial,georgiadis2023raster,nobari2017memory,xie2016simba}, we assume that $R$ and $S$ are memory-resident and static.
(The case of disk-resident data is beyond the scope of this paper.)
Each point $r_i \in R$ ($s_j \in S$) has a unique ID, i.e., $i$ ($j$), and is a 2-dimensional object, i.e., $r_i = \langle x_{r_i},y_{r_i}\rangle$, where $x_{r_i}$ and $y_{r_i}$ respectively represent the x and y coordinates of $r_i$.
For conciseness, we basically omit the ID of each point.

As a join predicate, we use an orthogonal range (or a window) because this shape is common and typical in spatial databases \cite{amagata2016monitoring,amagata2017general,ji2024safe,xie2021spatial,qi2018theoretically,qi2020packing,sowell2013experimental,pandey2018good,liu2021lhist} (and the other predicates are also not the scope of this paper).
Let $w(r)$ be an orthogonal rectangle centered at $r$.
If $s$ exists in $w(r)$, we use ``$w(r) \cap s$'' to represent it.
Then, the problem of spatial range join is defined as follows.

\begin{definition}[\textsc{Spatial range join}]
Given two point sets $R$ and $S$ and a range size, the spatial range join problem is to return
\begin{equation*}
    J = \{(r,s) \,|\, r \in R, s \in S, w(r) \cap s\}.
\end{equation*}
\end{definition}

\vs
\noindent
If $w(r) \cap s$, we have $r \cap w(s)$ since the range size is common for all points in $R$ and $S$.

As $|J|$ can be huge for large $n$ and $m$, we want to obtain only $t$ random pairs in $J$.
To obtain the distribution of $J$, these samples must be uniform and independent from any previous samples \cite{zhao2018random,deng2023join}.
The problem of random sampling over spatial range joins considers this requirement.

\begin{definition}[\textsc{Random sampling over spatial range join}]    \label{definition:problem}
Given two point sets $R$ and $S$, a range size, and the number of samples $t$, this problem returns $t$ samples of $J$, each of which is picked uniformly at random.
\end{definition}

\noindent
We address the problem in \cref{definition:problem}, i.e., random sampling with replacement over spatial range joins.
Even if random sampling without replacement is required, extending the algorithms introduced in this paper is straightforward (just rejecting a given sample if it has already been obtained).
Without loss of generality, we assume that $|J| \geq 1$.
Although \cref{definition:problem} uses $t$ as an input, it can be $\infty$.
Because all algorithms introduced in this paper pick join samples progressively, they can stop sampling whenever sufficient join samples are obtained.

This paper considers exact algorithms for this problem (i.e., algorithms that yield uniform and independent join samples).
The objective of this work is to design a theoretically- and empirically-efficient algorithm w.r.t. space and time.

\section{Baseline Algorithms}   \label{sec:baseline}
Before presenting our algorithm, we design two baseline algorithms that employ existing techniques.

\subsection{KDS}    \label{sec:baseline:kds}
First, we employ KDS \cite{xie2021spatial}, the latest algorithm that enables random sampling over a range search.
Given a point $r$ and $S$, KDS returns a random sample from $S(w(r))$, a subset of $S$ that overlaps $w(r)$, in $O(\sqrt{m})$ time by using a $k$d-tree.
KDS guarantees that samples are uniform and independent, because a point $s \in S(w(r))$ is sampled with probability $1 / |S(w(r))|$.

Here, it is important to note that
\begin{equation}
    J = \bigcup_{r \in R}\{(r,s)\,|\, s \in S(w(r))\}.
\end{equation}
Because $|S(w(r))|$ is different for each $r \in R$, sampling a point $r$ from $R$ uniformly at random and then running KDS with $w(r)$ does not satisfy uniform sampling probability.
We therefore use weighted sampling to satisfy uniform and independent spatial range join sampling.

\vs
\noindent
\textbf{Algorithm description.}
We describe the first baseline algorithm that employs KDS.
(We call this baseline algorithm KDS.)
A $k$d-tree of $S$ is built offline.
In a nutshell, KDS samples $r$ with weighted sampling based on $|S(w(r))|$, and then samples $s$ uniformly at random from $S(w(r))$, to sample $(r,s) \in J$ with probability $1/|J|$.

More specifically, given $R$, $S$, $t$, and a rectangle size,
\begin{enumerate}
    \setlength{\leftskip}{-3.0mm}
    \item   we run a range counting on the $k$d-tree, to obtain $|S(w(r))|$, for each $r \in R$.
    \item   We use Walker's alias method \cite{walker1974new} to enable efficient weighted sampling.
            This method builds a structure called \textit{alias} and returns an element with probability proportional to its weight.
            In our context, $r$ is picked with probability $\frac{|S(w(r))|}{\sum_{r' \in R}|S(w(r'))|}$.
    \item   Assume that the alias returns $r_{i}$.
            Then we run KDS with $w(r_{i})$ and obtain a random join sample $(r_{i}, s_{j})$.
    \item   We repeat the third step until we have $t$ join samples.
\end{enumerate}

\vs
\noindent
\textbf{Space complexity.}
The $k$d-tree needs $O(m)$ space, whereas the alias structure needs $O(n)$ space \cite{walker1974new}.
Therefore, KDS needs $O(n+m)$ space.

\vs
\noindent
\textbf{Time complexity.}
A range counting on a $k$d-tree of $m$ 2-dimensional points runs in $O(\sqrt{m})$ time \cite{de2000computational}.
Hence, the first step needs $O(n\sqrt{m})$ time.
The alias structure is built in $O(n)$ time \cite{walker1974new}, i.e., the second step needs $O(n)$ time.
This structure yields a weighted random sample in $O(1)$ time \cite{walker1974new}, so the third step needs $O(\sqrt{m})$ time.
In total, this baseline algorithm needs $O(n\sqrt{m}) + O(n) + t \times O(\sqrt{m}) = O((n+t)\sqrt{m})$ time.

\vs
\noindent
\textbf{Correctness.}
The alias picks $r$ with probability $\frac{|S(w(r))|}{\sum_{r' \in R}|S(w(r'))|} = \frac{|S(w(r))|}{|J|}$, and KDS picks $s$, such that $w(r) \cap s$, with probability $\frac{1}{|S(w(r))|}$.
That is, each pair in $J$ is sampled with probability $\frac{1}{|J|}$.
Samples obtained by the alias and KDS are not affected by previous samples, since they are picked uniformly at random.
Therefore, the correctness is obvious.

\subsection{KDS-rejection}  \label{sec:baseline:kds-r}
We next incorporate the rejection sampling approach employed in the state-of-the-art algorithm for join sampling in relational databases \cite{zhao2018random} into KDS.
The first baseline algorithm runs exact range counting to satisfy uniform and independent join sampling, and this approach incurs $O(n\sqrt{m})$ time.
This computational cost is huge when $n$ and $m$ are large.
The rejection sampling reduces this cost (but may increase the join sampling cost).
The main idea of the rejection sampling approach is to use $\mu(r)$, an upper-bound of $|S(w(r))|$, instead of $|S(w(r))|$ and reject a given sample with a certain probability.

Now assume that the size of a given orthogonal rectangle is $l \times l$.
(The height and width can be different sizes, but we use the same size for ease of explanation.)
For each point $s \in S$, we map it to a grid with non-empty cells, where each cell is a square with side length $\frac{l}{2}$.
Given a point $r \in R$, it is clear that $w(r)$ overlaps at most nine cells, see \cref{fig:grid}.
Let $c$ be a cell of the grid, and we use $S(c)$ to represent a set of points in $S$ that belong to $c$.
Then we see that $\mu(r)$ can be the sum of $|S(c)|$, where $c$ overlaps $w(r)$, and $\mu(r)$ can be obtained in $O(1)$ time.
We have $\mu(r) \geq |S(w(r))|$, because eight out of nine cells \textit{partially} overlap $w(r)$.

\vs
\noindent
\textbf{Algorithm description.}
Based on the above ideas, we design another baseline algorithm called KDS-rejection.
This algorithm also builds a $k$d-tree of $S$ offline, and this $k$d-tree is used for join sampling.
\begin{enumerate}
    \setlength{\leftskip}{-3.0mm}
    \item   We map each $s \in S$ to its corresponding cell.
            If this cell has not been created, we create it.
            This grid cannot be built offline, since the cell size depends on the rectangle size.
    \item   For each $r \in R$, we obtain $\mu(r)$ by accessing the cells overlapping $w(r)$.
    \item   As with the first baseline algorithm, we use Walker's alias method to build an alias.
            Note that this alias returns $r$ with probability proportional to $\mu(r)$.
    \item   Assume that the alias returns $r_{i}$.
            Then we run KDS with $w(r_{i})$ and obtain a random join sample $(r_{i}, s_{j})$ along with $|S(w(r))|$.
            We accept this join sample with probability $\frac{|S(w(r))|}{\mu(r)}$.
            (We reject it with probability $1 - \frac{|S(w(r))|}{\mu(r)}$.)
    \item   We repeat the fourth step until we have $t$ join samples.
\end{enumerate}

\vs
\noindent
\textbf{Space complexity.}
In addition to the $k$d-tree and alias, this algorithm uses a grid with at most $m$ cells, and each point $\in S$ belongs to a unique cell.
Hence, the space complexity is $O(n+m)$.

\vs
\noindent
\textbf{Time complexity.}
The first step (point mapping to grid) needs $O(m)$ time.
For a given point $r \in R$, obtaining $\mu(r)$ needs $O(1)$ time, so the second step needs $O(n)$ time.
The third and fourth steps respectively require $O(n)$ and $O(\sqrt{m})$ times, see \cref{sec:baseline:kds}.

We next focus on how many iterations are required to obtain $t$ join samples.
Obviously, we can obtain a join sample with probability $\frac{|J|}{\sum_{r \in R}\mu(r)}$.
The expected number of iterations is therefore $\frac{\sum_{r \in R}\mu(r)}{|J|}$.
Because $\mu(r)$ has no bound, we have $\mu(r) = O(m)$ in the worst case.
Summarizing the above facts, this algorithm needs $O(m) + O(n) + O(n) + t \times \frac{nm}{|J|} \times O(\sqrt{m}) = O(n+m+\frac{nm^{1.5}t}{|J|})$ time in expectation.

\vs
\noindent
\textbf{Correctness.}
The alias picks $r$ with probability $\frac{\mu(r)}{\sum_{r \in R}\mu(r)}$, and KDS picks $s$, such that $w(r) \cap s$, with probability $\frac{1}{|S(w(r))|}$.
This join sample $(r, s)$ is accepted with probability $\frac{|S(w(r))|}{\mu(r)}$.
Therefore, this pair is included in the result set with probability $\frac{1}{\sum_{r \in R}\mu(r)}$, satisfying the uniform and independent requirements.

\section{Our Algorithm} \label{sec:proposal}
Although the baseline algorithms satisfy the uniform and independent requirements and do not suffer from $O(nm)$ time, they are inefficient.
Their main drawbacks are twofold.
The first one is that KDS incurs a significant cost (i.e., $O(\sqrt{m})$ time) to pick a single join sample.
Therefore, both baseline algorithms cause a long delay when $m$ and/or $t$ are large.
The other drawback is their range counting.
KDS computes the exact range count (i.e., $|S(w(r))|$) for each $r \in R$ and incurs a significant computational cost, i.e., $O(n\sqrt{m})$ time.
KDS-rejection reduces this cost to $O(n)$ by computing an upper-bound of $|S(w(r))|$ for each $r \in R$.
However, this upper-bounding has no approximation bound, so its acceptance probability (in rejection sampling) can be low.
Such a low acceptance probability leads to many sampling iterations, which also causes a long delay.

These concerns pose the challenge of designing a data structure that can efficiently deal with both range counting and sampling.
We propose a new data structure that overcomes this challenge and guarantees a theoretical performance.

\subsection{Overview}   \label{sec:proposal:overview}
Our algorithm has three phases: online data structure building phase, approximate range counting phase, and sampling phase.
\cref{algo:proposed} describes the overview of our algorithm.
(Each of these phases will be elaborated on in the subsequent sections.)

\begin{figure}[!t]
    \centering
    \includegraphics[width=0.40\linewidth]{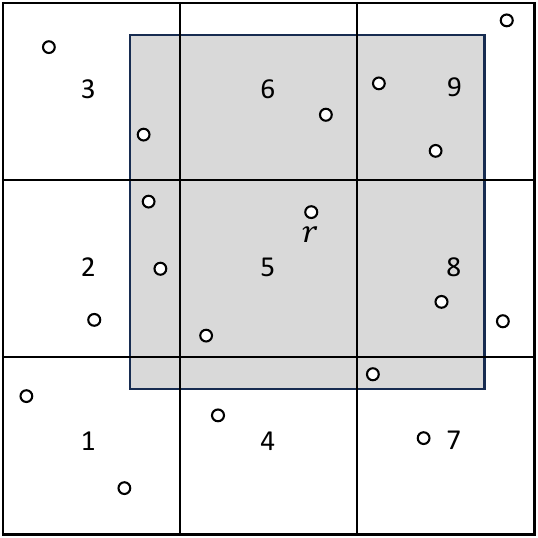}
    \vspace{-2.0mm}
    \caption{Grid and $w(r)$ (the grey rectangle)}
    \label{fig:grid}
    \vspace{-5.0mm}
\end{figure}

\vs
\noindent
$\triangleright$ \textbf{Online data structure building phase} (lines \ref{algo:proposed:1_b}--\ref{algo:proposed:1_e}).
We first map each point $s \in S$ to a grid $G$ with non-empty cells, as with KDS-rejection.
However, we use this grid with a different idea.
\cref{fig:grid} illustrates the cells that overlap $w(r)$ (the gray rectangle).
It is trivial that $w(r)$ overlaps at most nine cells.
These cells belong to one of the following three cases.
\begin{itemize}
    \setlength{\leftskip}{-2.0mm}
    \item   Case 1: the cell is fully covered (i.e., cell 5 in \cref{fig:grid}).
    \item   Case 2: the cell is fully covered w.r.t. the x- or y-dimension but partially covered w.r.t. the other dimension (i.e., cells 2, 4, 6, and 8 in \cref{fig:grid}).
    \item   Case 3: the cell is partially covered w.r.t. both the x- and y-dimensions (i.e., cells 1, 3, 7, and 9 in \cref{fig:grid}).
\end{itemize}
Without the grid, $w(r)$ is a 4-sided range.
On the other hand, with the grid, we can convert the 4-sided range to at most 2-sided range.
That is, case 1 is 0-sided, case 2 is 1-sided, and case 3 is 2-sided, which is easy to see from \cref{fig:grid}.

From this insight, we can easily and efficiently deal with cases 1 and 2 regarding range counting and sampling.
Case 1 allows $O(1)$ time exact range counting and $O(1)$ time random sampling.
Recall that $S(c)$ is a set of points in $S$ that belong to a cell $c$.
In case 2, we can run an exact range counting in $O(\log |S(c)|)$ time with a binary search if the points $\in S(c)$ are sorted based on the x- and y-dimensions.
Also, in this case, we can sample a point in $O(\log |S(c)|)$ time with a binary search.

The remaining challenge is to deal with case 3 efficiently.
To solve this challenge, we propose a new data structure, BBST (Bucket-based Binary Search Tree).
This structure enables $\tilde{O}(1)$-approximate range counting in $\tilde{O}(1)$ time.
To satisfy this result at an arbitrary cell, we need to build two BBSTs for each cell.
This building is done after grid mapping.

\begin{algorithm}[!t]
    \caption{Proposed Algorithm}	\label{algo:proposed}
    \DontPrintSemicolon
    \KwIn {$R$, $S$, $l$ (range), $t$}
    \KwOut{$t$ samples of $J$}
    {\small
    \tcc{{\footnotesize Online data structure building phase}}
    $G \gets$ \textsc{Grid-Mapping$(S,l)$}\;    \label{algo:proposed:1_b}
    \ForEach {cell $c \in G$}{
        $S^y(c) \gets S(c)$\;
        Sort the points in $S^y(c)$ in ascending order of the y-coordinate\;
        $\mathcal{T}^{min}_{c}$, $\mathcal{T}^{max}_{c} \gets$ \textsc{BBST-Building$(c)$}  \label{algo:proposed:1_e}
    }
    \vspace{1.0mm}
    \tcc{{\footnotesize Approximate range counting phase}}
    \ForEach {$r \in R$}{   \label{algo:proposed:2_b}
        $\mu(r) \gets$ \textsc{Upper-bounding$(r,G)$}\;
        $\mathcal{A}_{r} \gets$ \textsc{Alias-Building$(C_{r})$}  \Comment*[r]{\scriptsize $C_{r}$ is a set of cells overlapping $w(r)$}
    }
    $\mathcal{A} \gets$ \textsc{Alias-Building$(\{\mu(r_{1}), ..., \mu(r_{n})\})$}\;  \label{algo:proposed:2_e}
    \vspace{1.0mm}
    \tcc{{\footnotesize Sampling phase}}
    $T \gets \varnothing$\; \label{algo:proposed:3_b}
    \While {$|T| < t$}{
        $r \gets$ \textsc{Weighted-Sampling$(A)$}\;     \label{algo:proposed:wsample-1}
        $c \gets$ \textsc{Weighted-Sampling$(A_r)$}\;   \label{algo:proposed:wsample-2}
        $s \gets$ \textsc{Sampling$(c, r, l)$}    \Comment*[r]{\scriptsize sample a point $\in S$ from $c$ or $\mathcal{B}_{c}$}
        \textbf{If} $w(r) \cap s$ \textbf{then} $T \gets T \cup \{(r, s)\}$ \label{algo:proposed:3_e}
    }
    \textbf{return} $T$
    }
\end{algorithm}

\vs
\noindent
$\triangleright$ \textbf{Approximate range counting phase} (lines \ref{algo:proposed:2_b}--\ref{algo:proposed:2_e}).
For each point $r \in R$, we compute $\mu(r)$, an upper-bound of $|S(w(r))|$, with a new approximate range counting.
We exploit the grid to compute $\mu(r)$.
Specifically, we compute the exact range counts for cases 1 and 2 (different from KDS-rejection).
On the other hand, for case 3, we obtain $\tilde{O}(1)$-approximate range counts with the BBSTs.
Thanks to this approach, we have $\mu(r) \leq \tilde{O}(|S(w(r))|)$, and this is obtained in only $\tilde{O}(1)$ time.
From $\{\mu(r_1), ..., \mu(r_n)\}$, we build an alias structure, as with the baseline algorithms.

\vs
\noindent
$\triangleright$ \textbf{Sampling phase} (lines \ref{algo:proposed:3_b}--\ref{algo:proposed:3_e}).
We pick a join sample so that this sample is independent and uniform.
To satisfy this efficiently, we use the grid structure and BBSTs.
In any case, we can pick a join sample in at most $\tilde{O}(1)$ time.
We repeat this until $t$ join samples are obtained.

\subsection{Bucket-based Binary Search Tree}    \label{sec:proposal:bbst}
Before presenting the three steps in detail, we introduce the new data structure BBST.

\vs
\noindent
\textbf{Main ideas.}
Recall that this data structure is designed to deal with 2-sided rectangles, i.e., case 3 in \cref{sec:proposal:overview}.
Our first idea here is to enable binary searches in both the x- and y-dimensions (e.g., a binary search on the x-dimension and then another one on the y-dimension).
However, simply implementing this idea on a set $S(c)$ of points in a given cell $c \in G$ needs a space larger than $O(|S(c)|)$ \cite{bentley1979decomposable}.
This means that the space complexity of this approach cannot be linear to $m$, whereas linear space complexity is required to scale to a large number of points.

To remove this issue, we bundle some points into a bucket.
We then build a balanced binary search tree on a set of such buckets based on the x-dimension.
To enable binary searches based on the y-dimension, we make each node store some sorted arrays of buckets existing in the sub-tree rooted at this node.

\vs
\noindent
\textbf{Data structure.}
Since a BBST is built from a set of buckets, we first define \textit{bucket}.

\begin{definition}[\textsc{Bucket}] \label{definition:bucket}
Given a set $S(c)$ of points belonging to a cell $c$, a bucket $B \subseteq S(c)$ is a sequence of $\log m$ points sorted in ascending order of the x-coordinate.
Additionally, we record $\min_{s \in B}s_{x}$, $\max_{s \in B}s_{x}$, $\min_{s \in B}s_{y}$, and $\max_{s \in B}s_{y}$.
\end{definition}

\begin{figure}[!t]
    \centering
    \includegraphics[width=0.49\linewidth]{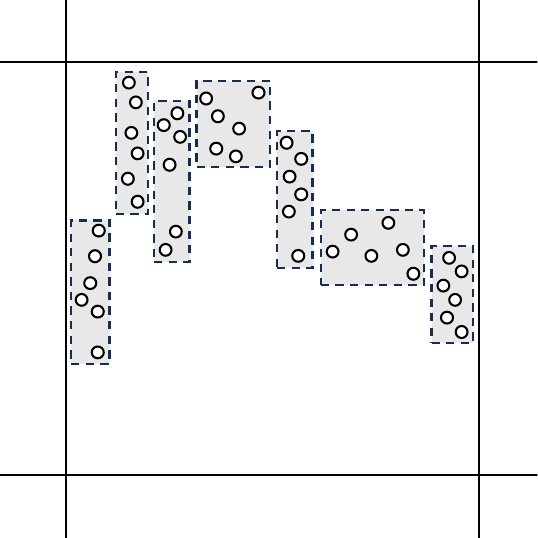}
    \caption{Buckets in a cell}
    \label{fig:bucket}
    \vspace{-5.0mm}
\end{figure}

\noindent
\cref{fig:bucket} illustrates buckets (each of which is a set of points in the same gray rectangle) in a cell.
The size of $B$ is carefully set to have the performance guarantee (see \cref{lemma:bound} and \cref{theorem:time}.)

The base structure of a BBST $\mathcal{T}_c$ of $S(c)$ is a balanced binary search tree.
For now, we use $\min_{s \in B}s_{x}$ as the x-coordinate of $B$ for presentation.
With this assumption, $\mathcal{T}_c$ is built based on the x-dimension.
Each node $u_i$ of $\mathcal{T}_c$ has the following elements:
\begin{itemize}
    \setlength{\leftskip}{-2.0mm}
    \item   two lists $\mathcal{B}_{i}^{min}$ and $\mathcal{B}_{i}^{max}$ of buckets that have the same x-coordinate with $u_i$.
            The buckets in $\mathcal{B}_{i}^{min}$ ($\mathcal{B}_{i}^{max}$) are sorted in ascending order of $\min_{s \in B}s_{y}$ ($\max_{s \in B}s_{y}$).
    \item   two sorted arrays $A_{i}^{min}$ and $A_{i}^{max}$ of buckets existing in the sub-tree rooted at $u_i$.
            The buckets in $A_{i}^{min}$ ($A_{i}^{max}$) are sorted in ascending order of $\min_{s \in B}s_{y}$ ($\max_{s \in B}s_{y}$).
    \item   a pointer to its left child node and a pointer to its right child node. 
\end{itemize}

Notice that the first two elements differentiate BBST from a standard binary search tree.
The first element is important to guarantee that the BBST is balanced even if we have buckets (i.e., points) with the same x-coordinates.
The second element is used for binary searches w.r.t. the y-dimension.

\vs
\noindent
\textbf{Building algorithm.}
Given a set $S(c)$ of points in a cell $c$, we partition it into buckets by following \cref{definition:bucket} and obtain a set $\mathcal{B}$ of buckets, as shown in \cref{fig:bucket}.
Then, we make two copies of $\mathcal{B}$.
One is sorted in ascending order of $\min_{s \in B}s_{y}$, and the other is sorted in ascending order of $\max_{s \in B}s_{y}$.
We next compute the median of the x-coordinates in $\mathcal{B}$, create a root node $u_{root}$ of $\mathcal{T}_c$, and compute $\mathcal{B}_{root}^{min}$, $\mathcal{B}_{root}^{max}$, $A_{root}^{min}$, and $A_{root}^{max}$.
After that, based on the x-coordinate of $u_{root}$, we partition $\mathcal{B}$ and its two copies for the left and right child nodes of $u_{root}$.
These nodes and their associated structures are created similarly, and $\mathcal{T}_c$ is built in this recursive manner.

Recall that we need to create two BBSTs in a cell.
In the above presentation, for each bucket $B \in \mathcal{B}$, we used $\min_{s \in B}s_{x}$ as the x-coordinate of $B$.
We build another BBST of $\mathcal{B}$ by using $\max_{s \in B}s_{x}$ as the x-coordinate of $B \in \mathcal{B}$.
\cref{algo:build} details the algorithm for building two BBSTs.

\begin{algorithm}[!t]
    \caption{\textsc{BBST-Building}$(c)$}	\label{algo:build}
    \DontPrintSemicolon
    \KwIn {$c$}
    {\small
    Partition $S(c)$ into buckets $B_1, \cdots, B_{b}$, each of which has (at most) $\log m$ points \Comment*[r]{\scriptsize points in $S(c)$ are pre-sorted based on the x-coordinate}
    $\mathcal{B}, \mathcal{B}_{cp1}, \mathcal{B}_{cp2} \gets$ a set of buckets $B_1, \cdots, B_{b}$\;
    Sort the buckets in $\mathcal{B}_{cp1}$ ($\mathcal{B}_{cp2}$) in ascending order of $\min_{s \in B \in \mathcal{B}}s_{y}$ ($\max_{s \in B \in \mathcal{B}}s_{y}$)\;
    root node of $\mathcal{T}_{c}^{min} \gets$ \textsc{Make-Node}$(\mathcal{B}, \mathcal{B}_{cp1}, \mathcal{B}_{cp2}, 1)$\;
    root node of $\mathcal{T}_{c}^{max} \gets$ \textsc{Make-Node}$(\mathcal{B}, \mathcal{B}_{cp1}, \mathcal{B}_{cp2}, 0)$\;
    \SetKwProg{Fn}{Function}{:}{}
    \Fn{\textsc{Make-Node}$(\mathcal{B}, \mathcal{B}_{cp1}, \mathcal{B}_{cp2}, f)$}{
        \tcc{{\footnotesize If $f = 1$ ($f = 0$), $B.x = \min_{s \in B}s_{x}$ ($\max_{s \in B}s_{x}$)}}
        \textbf{if} $\mathcal{B} = \varnothing$ \textbf{then return}\;
        $u_{i} \gets$ a new node\;
        $u_{i}.x \gets$ median among $\{B.x \,|\, B \in \mathcal{B}\}$\;
        $\mathcal{B}_{l}, \mathcal{B}_{r}, \mathcal{B}_{cp1, l}, \mathcal{B}_{cp1, r}, \mathcal{B}_{cp2, l}, \mathcal{B}_{cp2, r} \gets \varnothing$\;
        $A_{i}^{min} \gets \mathcal{B}_{cp1}$, $A_{i}^{max} \gets \mathcal{B}_{cp2}$\;
        \ForEach{$B \in \mathcal{B}$}{
            \textbf{if} $B.x < u_{i}.x$ \textbf{then} Add $B$ into $\mathcal{B}_{l}$\;
            \textbf{else if} $B.x > u_{i}.x$ \textbf{then} Add $B$ into $\mathcal{B}_{r}$
        }
        \ForEach{$B \in \mathcal{B}_{cp1} (\mathcal{B}_{cp2})$}{
            \uIf {$B.x = u_{i}.x$}{
                Add $B$ into $\mathcal{B}_{i}^{min}$ ($\mathcal{B}_{i}^{max}$)
            }
            \uElseIf {$B.x < u_{i}.x$}{
                Add $B$ into $\mathcal{B}_{cp1, l}$ ($\mathcal{B}_{cp2, l}$)
            }
            \Else {
                Add $B$ into $\mathcal{B}_{cp1, r}$ ($\mathcal{B}_{cp2, r}$)
            }
        }
        \textbf{if} $|\mathcal{B}| = 1$ \textbf{then return} $u_i$\;
        left child of $u_i \gets$ \textsc{Make-Node}$(\mathcal{B}_l, \mathcal{B}_{cp1, l}, \mathcal{B}_{cp2, l}, f)$\;
        right child of $u_i \gets$ \textsc{Make-Node}$(\mathcal{B}_r, \mathcal{B}_{cp1, r}, \mathcal{B}_{cp2, r}, f)$
        }
    }
\end{algorithm}

\vs
\noindent
\textbf{Building time} of two BBSTs in a cell is reasonable.

\begin{lemma}   \label{lemma:bbst:build-time}
Given a set of $N \leq m$ points, the time complexity of building its BBST is $O(N)$.
\end{lemma}

\noindent
\textsc{Proof.}
We assume that points in $S$ are sorted in ascending order of the x-coordinate, which can be done in a pre-processing step\footnote{This sorting is not necessarily done offline.
Even if we do it online, our theoretical time does not change.}.
We can make buckets in $O(N)$ time, and we have $O(\frac{N}{\log m})$ buckets.
Two copies of $\mathcal{B}$ are sorted based on the y-dimension, which is done in $O(\frac{N}{\log m}\log \frac{N}{\log m}) = O(\frac{N}{\log m}\log N) \leq O(N)$ time.
When we create a node, we compute the median of x-coordinates and scan $\mathcal{B}$ and its two copies, and these are done in $O(\frac{N}{\log m})$ time.
Because the height of a BBST is $O(\log \frac{N}{\log m}) = O(\log N)$, it is easy to see that a bucket is accessed at most $O(\log N)$ times.
Therefore, the total cost of building a BBST of $N$ points is $O(N)$.
\wsq

\vs
\noindent
This lemma clarifies that two (i.e., a constant number of) BBSTs are built in $O(N)$ for a cell containing $N$ points.

\vs
\noindent
\textbf{Space complexity.}
We have the following result w.r.t. the space complexity of a BBST.
This result clarifies that the space of our data structure is only linear to the number of points. 

\begin{lemma}   \label{lemma:bbst:space}
A BBST of a set of $N \leq m$ points needs $O(N)$ space.    
\end{lemma}

\noindent
\textsc{Proof.}
Since the base structure of the BBST is a balanced binary search tree, it has at most $O(\frac{N}{\log m})$ nodes.
Now consider $A_{i}^{min}$ of a node $u_{i}$, and the size of $A_{i}^{min}$ is equal to the number of nodes existing in the sub-tree rooted at $u_{i}$.
This means that a bucket is stored in at most $O(\log\frac{N}{\log m})$ nodes.
The total space is hence $O(\frac{N}{\log m}) \times O(\log\frac{N}{\log m}) \leq O(N)$ since $N \leq m$.
\wsq

\subsection{Online Data Structure Building Phase}
\noindent
\textbf{Algorithm description.}
This phase first maps each point $s \in S$ into its corresponding cell.
When we do not have such a cell for $s$, we create this cell, resulting in a non-empty grid $G$.
This procedure corresponds to \textsc{Grid-Mapping$(S,l)$} in \cref{algo:proposed}.
Then, for each cell $c \in G$, we make a copy of $S(c)$, denoted by $S^{y}(c)$, and we sort the points in $S^{y}(c)$ based on the y-coordinate.
The points in $S(c)$ are sorted based on the x-coordinate, since the points $\in S$ are pre-sorted based on the x-dimension.
We build two BBSTs with \cref{algo:build}.

\vs
\noindent
\textbf{Time complexity in this phase} is analyzed below:

\begin{lemma}   \label{lemma:phase1-time}
The online data structure building phase needs $O(m\log m)$ time.
\end{lemma}

\noindent
\textsc{Proof.}
Because mapping a point $s$ into its corresponding cell needs $O(1)$ time, \textsc{Grid-Mapping$(S,l)$} needs $O(m)$ time.
Given a cell $c$, sorting the points in $S^{y}(c)$ needs $O(|S^{y}(c)|\log |S^{y}(c)|)$ time.
Also, from \cref{lemma:bbst:build-time}, \textsc{BBST-Building}$(c)$ needs $O(|S(c)|)$ time.
In total, we need $\sum_{c \in G}O(|S^{y}(c)|\log |S^{y}(c)|) \leq O(m\log m)$ time.
\wsq

\begin{figure*}[!t]
    \centering
    \includegraphics[width=0.999\linewidth]{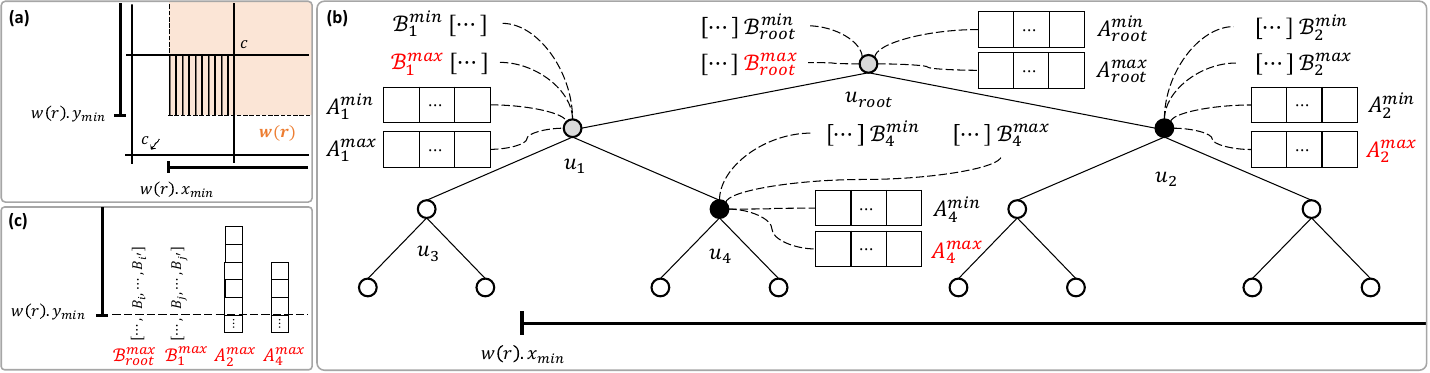}
    \vspace{-5.0mm}
    \caption{Example of case 3 ($c_{\swarrow}$).
    (a) Given $w(r)$, we approximately count the number of points in the striped space of $c_{\swarrow}$. The search space is $[w(r).x_{min},\infty) \times [w(r).y_{min},\infty)$.
    (b) Example of traversing $\mathcal{T}^{max}_{c_{\swarrow}}$. Black nodes are canonical nodes, whereas gray nodes also satisfy the query range $[w(r).x_{min}, \infty)$.
    The associated structures, which are used to compute $\mu(r,c_{\swarrow})$, are marked by red.
    (c) After traversing $\mathcal{T}^{max}_{c_{\swarrow}}$, with the query point of $w(r).y_{min}$, we obtain $\mu(r,c_{\swarrow})$ by counting the number of buckets $B$, such that $w(r).y_{min} \leq \max_{s\in B}s.y$, in the lists and arrays identified from $\mathcal{T}^{max}_{c_{\swarrow}}$.}
    \label{fig:canonical}
    \vspace{-4.0mm}
\end{figure*}

\subsection{Approximate Range Counting Phase}
Recall that $S(w(r))$ is a set of points $\in S$ overlapping $w(r)$.
In this phase, we compute an upper-bound $\mu(r)$ of $|S(w(r))|$ for each $r \in R$.
Our idea of facilitating upper-bound computation is to decompose $w(r)$ into nine rectangles with the cells overlapping $w(r)$, as shown in \cref{fig:grid}.
Let $c$ be the cell where $r$ belongs.
Furthermore, let $c_{\swarrow}$ be the left-bottom cell of $c$ (e.g., cell 1 when $c$ is cell 5 in \cref{fig:grid}).
In this sense, we use arrows to represent the neighbor cells of $c$.

\vs
\noindent
\textbf{Rationale of computing $\mu(r)$.}
As introduced in \cref{sec:proposal:overview}, we classify each cell based on the number of sides held by $w(r)$ w.r.t. the overlapped rectangle between $w(r)$ and a cell.
Let $\mu(r,c)$ be an upper-bound of the number of points in $w(r)$ and $c$.
Trivially, $\mu(r) = \mu(r,c_{\swarrow}) + \mu(r,c_{\leftarrow}) + \mu(r,c_{\nwarrow}) + \mu(r,c_{\downarrow}) + \mu(r,c) + \mu(r,c_{\uparrow}) + \mu(r,c_{\searrow}) + \mu(r,c_{\rightarrow}) + \mu(r,c_{\nearrow})$.

\vs
\noindent
\underline{(i) \textit{Case} 1 (0-sided: $c$).}
Since $c$ is fully covered by $w(r)$,
\begin{equation*}
    \mu(r,c) = |S(c)|.
\end{equation*}

\vs
\noindent
\underline{(ii) \textit{Case} 2 (1-sided: $c_{\leftarrow}$, $c_{\downarrow}$, $c_{\uparrow}$, and $c_{\rightarrow}$).}
Let $w(r).x_{min}$ ($w(r).x_{max}$) be the left (right) x-coordinate of $w(r)$.
Also, let $w(r).y_{min}$ ($w(r).y_{max}$) be the bottom (top) y-coordinate of $w(r)$.
In the case of $c_{\leftarrow}$, the y-coordinates of the points in $S(c_{\leftarrow})$ are definitely within $w(r)$.
Therefore, we have
\begin{equation*}
    \mu(r,c_{\leftarrow}) = |\{s \,|\, s \in S(c_{\leftarrow}), w(r).x_{min} \leq s_{x}\}|.
\end{equation*}
Similarly,
\begin{align*}
    \mu(r,c_{\rightarrow}) &= |\{s \,|\, s \in S(c_{\rightarrow}), s_{x} \leq w(r).x_{max}\}|,   \\
    \mu(r,c_{\downarrow}) &= |\{s \,|\, s \in S(c_{\downarrow}), w(r).y_{min} \leq s_{y}\}|, \,\text{and} \\
    \mu(r,c_{\uparrow}) &= |\{s \,|\, s \in S(c_{\uparrow}), s_{y} \leq w(r).y_{max}\}|.
\end{align*}

\vs
\noindent
\underline{(iii) \textit{Case} 3 (2-sided: $c_{\swarrow}$, $c_{\nwarrow}$, $c_{\searrow}$, and $c_{\nearrow}$).}
We use the case of $c_{\swarrow}$ for explanation.
To compute $\mu(r,c_{\swarrow})$, we search for the buckets that overlap $w(r)$.
Consider a range of $[w(r).x_{min}, c_{\swarrow}.x_{max}]$, where $c_{\swarrow}.x_{max}$ is the right x-coordinate of $c_{\swarrow}$.
Each bucket $B$ in $c_{\swarrow}$ has $\max_{s \in B}s_{x} \leq c_{\swarrow}.x_{max}$, so it may overlap $w(r)$ if $w(r).x_{min} \leq \max_{s \in B}s_{x}$.
Also, for a bucket $B$ in $c_{\swarrow}$, if $\max_{s \in B}s_{y} \geq w(r).y_{min}$, $B$ may overlap $w(r)$.
These facts and \cref{definition:bucket} derive
\begin{align*}
    &\mu(r,c_{\swarrow}) = \log m \,\times\\
    &|\{B \,|\, B \in \mathcal{B}(c_{\swarrow}), w(r).x_{min} \leq \max_{s \in B}s_{x}, w(r).y_{min} \leq \max_{s \in B}s_{y}\}|,
\end{align*}
where $\mathcal{B}(c_{\swarrow})$ is a set of buckets in $c_{\swarrow}$.
Similarly,
\begin{align*}
    &\mu(r,c_{\nwarrow}) = \log m \,\times\\
    &|\{B \,|\, B \in \mathcal{B}(c_{\nwarrow}), w(r).x_{min} \leq \max_{s \in B}s_{x}, \min_{s \in B}s_{y} \leq w(r).y_{max}\}|,   \\
    &\mu(r,c_{\searrow}) = \log m \,\times\\
    &|\{B \,|\, B \in \mathcal{B}(c_{\searrow}), \min_{s \in B}s_{x} \leq w(r).x_{max}, w(r).y_{min} \leq \max_{s \in B}s_{y}\}|,
\end{align*}
and
\begin{align*}
    &\mu(r,c_{\nearrow}) = \log m \,\times\\
    &|\{B \,|\, B \in \mathcal{B}(c_{\nearrow}),  \min_{s \in B}s_{x} \leq w(r).x_{max}, \min_{s \in B}s_{y} \leq w(r).y_{max}\}|.
\end{align*}

\noindent
\textbf{Algorithm description.}
\textsc{Upper-bounding$(r,G)$} is described in \cref{algo:proposed}, and it is designed based on the above rationale.
Given a point $r \in R$, we access $c$ and its neighbor cells.
The following approach addresses each of these cells.

\vs
\noindent
\underline{(i) \textit{Case} 1.}
We return the size of $S(c)$.

\vs
\noindent
\underline{(ii) \textit{Case} 2.}
This case is also easy, thanks to the grid structure.
In the case of $c_{\leftarrow}$, we run a binary search on $S(c_{\leftarrow})$ with $w(r).x_{min}$.
From the position obtained by this binary search, it is clear that $\mu(r,c_{\leftarrow})$ is obtained.
The cases of $c_{\rightarrow}$, $c_{\downarrow}$, and $c_{\uparrow}$ are handled similarly.

\vs
\noindent
\underline{(iii) \textit{Case} 3.}
This is the most challenging case among the three cases.
We solve this challenge by using a BBST.

In the case of $c_{\swarrow}$, we use $\mathcal{T}^{max}_{c_{\swarrow}}$ since a bucket $B$ needs to satisfy $w(r).x_{min} \leq \max_{s \in B}s_{x}$ to overlap $w(r)$.
We traverse $\mathcal{T}^{max}_{c_{\swarrow}}$ from its root node $u_{root}$ with a query of $w(r).x_{min}$.
If $u_{root}.x < w(r).x_{min}$, we traverse its right child node.
Otherwise, we see that its right sub-tree is fully covered by $[w(r).x_{min}, \infty)$ w.r.t. the x-dimension.
Hence, we insert this canonical node (e.g., the right child node of $u_{root}$ in this case) into a set $U$.
In addition, $u_{root}$ is inserted into $U$ as the x-coordinates of the buckets in $\mathcal{B}_{root}$ are larger than or equal to $w(r).x_{min}$.
We then traverse the left child node of $u_{root}$.
(If $u_{root}.x = w(r).x_{min}$, we terminate the traversal.)
We traverse $\mathcal{T}^{max}_{c_{\swarrow}}$ in this recursive manner.

After we obtain $U$, we compute $\mu(r,c_{\swarrow})$ by using $A^{max}$ and $\mathcal{B}^{max}$.
For each $u_{i} \in U$, we run a binary search on $A^{max}_{i}$ or $\mathcal{B}^{max}_{i}$ with $w(r).y_{min}$ (or use the fractional cascading technique \cite{chazelle1986fractional}).
That is, given a node $u_{i} \in U$, we obtain
\begin{align}   \label{eq:bound}
    &\mu(r,c_{\swarrow},u_i) = \log m \,\times  \nonumber   \\
    &\begin{cases}
        |\{B \,|\, B \in A^{max}_{i},\, \theta(B,c_{\swarrow}) = 1\}|           & (\text{$u_i$ is a canonical node})\\
        |\{B \,|\, B \in \mathcal{B}^{max}_{i},\, \theta(B,c_{\swarrow}) = 1\}| & (\text{otherwise}),
    \end{cases}
\end{align}
where $\theta(B,c_{\swarrow})$ is true iff $w(r).x_{min} \leq \max_{s \in B}s_{x}$ and $w(r).y_{min} \leq \max_{s \in B}s_{y}$.
Then, it is clear that
\begin{equation*}
    \mu(r,c_{\swarrow}) = \sum_{u_i \in U}\mu(r,c_{\swarrow},u_i).
\end{equation*}

\begin{example}
\cref{fig:canonical} illustrates an example of our algorithm for case 3 ($c_{\swarrow}$).
See \cref{fig:canonical}(a), and we approximately count the number of points in the striped space of $c_{\swarrow}$.
For $c_{\swarrow}$, the query range is $[w(r).x_{min},\infty) \times [w(r).y_{min},\infty)$ since, as mentioned in \Cref{sec:proposal:overview}, it considers 2-sided range.

We first consider the x-dimension, i.e., $[w(r).x_{min},\infty)$, which is illustrated at the bottom of \cref{fig:canonical}(b).
With the BBST $\mathcal{T}^{max}_{c_{\swarrow}}$, we identify the lists/arrays containing the buckets $B$ such that $w(r).x_{min} \leq \max_{s \in B}s.x$.
Recall that each node is associated with the x-dimension.
Because $w(r).x_{min} \leq u_{root}.x$, we insert $u_{root}$ and $u_{2}$ into $U$.
(Notice that $u_2$ is a canonical node.)
Then, we traverse its left child node $u_1$ and do similar operations, resulting in $U = \{u_{root}, u_1, u_2, u_4\}$.

We next consider the y-dimension, i.e., $[w(r).y_{min},\infty)$, and use $\mathcal{B}^{max}_{root}$, $\mathcal{B}^{max}_{1}$, $A^{max}_{2}$, and $A^{max}_{4}$.
(Recall that $u_2$ and $u_4$ are canonical nodes, and every bucket $B$ in these lists and arrays satisfies that $w(r).x_{min} \leq \max_{s \in B}s.x$.)
As shown in \cref{fig:canonical}(c), we obtain $\mu(r,c_{\swarrow})$ by counting the number of buckets $B'$ such that $w(r).y_{min} \leq \max_{s \in B'}s.y$ with a query point of $w(r).y_{min}$.
\end{example}

The cases of $c_{\nwarrow}$, $c_{\searrow}$, and $c_{\nearrow}$ are handled similarly, so we note the differences from the case of $c_{\swarrow}$.
\begin{itemize}
    \item   The $c_{\nwarrow}$ case employs $\mathcal{T}^{max}_{c_{\nwarrow}}$.
            In the y-dimension, we employ $A^{min}$ and $\mathcal{B}^{min}$ with a query of $w(r).y_{max}$.
    \item   The $c_{\searrow}$ case employs $\mathcal{T}^{min}_{c_{\searrow}}$ and uses $w(r).x_{max}$ as a query in the x-dimension.
            In the y-dimension, we follow the case of $c_{\swarrow}$.
    \item   The $c_{\nearrow}$ case employs $\mathcal{T}^{min}_{c_{\nearrow}}$ and uses $w(r).x_{max}$ as a query in the x-dimension.
            In the y-dimension, we follow the case of $c_{\nwarrow}$.
\end{itemize}

After obtaining $\mu(r)$, we build an alias structure $\mathcal{A}_r$ so that, for $r$, we can pick a cell $\in \{c, c_{\swarrow}, ..., c_{\nearrow}\}$ with probability proportional to $\mu(r,c_{\mathchar`-})$, where $c_{\mathchar`-}$ is $c$, $c_{\swarrow}$, ..., or $c_{\nearrow}$.
In addition, after obtaining $\mu(r)$ for each $r \in R$, we build an alias structure $\mathcal{A}$ so that we can pick $r$ with probability $\frac{\mu(r)}{\sum_{r' \in R}\mu(r')}$.

\vs
\noindent
\textbf{Time complexity in this phase} is seen from:

\begin{lemma}   \label{lemma:counting-time}
The approximate range counting phase needs $O(n\log m)$ time.
\end{lemma}

\noindent
\textsc{Proof.}
We use a point $r \in R$ to prove this lemma.
Case 1 has only a single cell, and we need $O(1)$ time to obtain $\mu(r,c)$ clearly.
Case 2 has at most $4 = O(1)$ cells, and a binary search for obtaining $\mu(r,c_{\leftarrow})$ needs $O(\log |S(c_{\leftarrow})|) \leq O(\log m)$ time.
Therefore, case 2 needs $O(\log m)$ time.
Case 3 also has $4 = O(1)$ cells at most.
We traverse $O(\log N)$ nodes of a BBST of $N$ points, so we have $|U| = O(\log m)$.
For each $u_{i} \in U$, we run a binary search on $A^{max}_{i}$ or $\mathcal{B}^{max}_{i}$.
Because we have $|A^{max}_{i}| = O(\frac{N}{\log m})$ and $|\mathcal{B}^{max}_{i}| = O(\frac{N}{\log m})$, computing $\mu(r,c_{\mathchar`-})$ needs $O(\log m) \times O(\log m) = O(\log^{2}m)$ time.
Fractional cascading can reduce this cost to $O(\log m)$.
As a result, case 3 takes $O(\log m)$ time.
Building $\mathcal{A}_r$ needs $O(1)$ time, whereas building $\mathcal{A}$ needs $O(n)$ time.
Consequently, this phase needs $n \times O(\log m) + O(n) = O(n\log m)$ time.
\wsq

\vs
\noindent
\textbf{Approximation bound.}
We next prove the following lemma.

\begin{lemma}   \label{lemma:bound}
For each $r \in R$, we have
\begin{equation*}
    |S(w(r))| \leq \mu(r) \leq \max \{O(\log m) \times |S(w(r))|, O(\log m)\}.
\end{equation*}
\end{lemma}

\vs
\noindent
\textsc{Proof.}
Given an arbitrary point $r \in R$, it is trivial that cases 1 and 2 return the exact number of points in $w(r)$ and $c'$, where $c'$ falls into case 1 or 2.
Therefore, we focus on case 3.
Since case 3 finds buckets that overlap $w(r)$, it is guaranteed that $|S(w(r))| \leq \mu(r)$, thus we aim at proving $\mu(r) \leq \max \{O(\log m) \times |S(w(r))|, O(\log m)\}$.
To analyze this bound, we use the case of $c_{\swarrow}$ because the other cases have essentially the same analysis.

Consider a bucket $B$ that satisfies \cref{eq:bound}.
It is important to note that the set described in \cref{eq:bound} has no duplication.
That is, for arbitrary $A_{i}^{max}$, $A_{j}^{max}$, and $B_{k}^{max}$ such that $u_i$ and $u_j$ are canonical nodes in $U$ while $u_k \in U$ is not a canonical node, we have $A_{i}^{max} \cap A_{j}^{max} = \varnothing$ and $A_{i}^{max} \cap B_{j}^{max} = \varnothing$.
This bucket $B$ is classified into two sub-cases: (a) $w(r).x_{min} \leq \min_{s \in B}s.x$ and (b) $\min_{s \in B}s.x \leq w(r).x_{min}$.
In the former sub-case, there is at least one point $s \in B$ such that $w(r) \cap s$.
Therefore, the worst bound in this sub-case is clearly ``$\times$'' $\log m$.
On the other hand, in the latter sub-case, there may be no points $\in B$ that exist in $w(r)$, so the bound in this sub-case is ``$+$'' $\log m$.
From \cref{definition:bucket}, we have at most one bucket that falls into the latter sub-case.

Let $\alpha = \frac{\mu(r,c_{\swarrow})}{\log m}$, and $\alpha$ suggests the number of buckets overlapping $w(r)$.
When $\alpha \geq 2$, at least $\alpha - 1$ points exist in $w(r)$, so the relative error is at most $\frac{\alpha\log m}{\alpha - 1} = O(\log m)$.
When $\alpha = 1$, $\mu(r,c_{\swarrow}) \leq \log m$.
\cref{lemma:bound} therefore holds.
\wsq

\vs
As \cref{sec:experiment:result} introduces, the error of our approximate range counting is much better than $O(\log m)$ in practice.
This is because five out of nine cells yield exact range counts and the other cells (i.e., case 3) rarely return the worst bound in \cref{lemma:bound}.

\subsection{Sampling Phase} \label{sec:proposal:sampling}
\noindent
\textbf{Rationale.}
Given a point $r \in R$, to search for a random sample $s \in S$ such that $w(r) \cap s$, we exploit the search space used in the approximate range counting phase.
It is important to notice that \textit{our approximate range counting algorithm identifies the spaces where the points or buckets overlapping $w(r)$ exist}.
From this idea, we re-use (and extend) this algorithm to obtain spatial range join samples uniformly and independently.

\vs
\noindent
\textbf{Algorithm description.}
The following operations are iterated until we have $t$ join samples.
First, we use the alias $\mathcal{A}$ to randomly pick a point $\in R$ (line \ref{algo:proposed:wsample-1} of \cref{algo:proposed}).
Assume that we sample a point $r$.
We next want to sample a point $\in S$ such that $w(r) \cap s$.
We use the alias $\mathcal{A}_{r}$ to pick a cell that overlaps $w(r)$ (line \ref{algo:proposed:wsample-2} of \cref{algo:proposed}).
Assume that we pick a cell $c'$.
This cell, again, falls into one of the three cases.

\vs
\noindent
\underline{(i) \textit{Case} 1 ($c' = c$).}
We randomly pick a point $s \in S(c')$.

\vs
\noindent
\underline{(ii) \textit{Case} 2 ($c' = c_{\leftarrow}$, $c_{\downarrow}$, $c_{\uparrow}$, or $c_{\rightarrow}$).}
Our approach is the same as that for range counting.
For example, assume $c' = c_{\leftarrow}$.
We run a binary search on $S(c_{\leftarrow})$ with $w(r).x_{min}$ to identify the ``sequence'' of points $\in S(c_{\leftarrow})$ that overlap $w(r)$.
Then, we randomly pick a point from this sequence.
When $c' = c_{\downarrow}$, $c_{\uparrow}$, or $c_{\rightarrow}$, this approach is employed as well.

\vs
\noindent
\underline{(iii) \textit{Case} 3 ($c' = c_{\swarrow}$, $c_{\nwarrow}$, $c_{\searrow}$, or $c_{\nearrow}$).}
This case also employs an approach that is essentially similar to that in approximate range counting, but we have one main difference.
We \textit{probabilistically} determine whether we pick a sample at a given node.
Let us assume $c' = c_{\swarrow}$.

Set $\mu = \mu(r, c_{\swarrow})$.
At a root node $u_{root}$ of $\mathcal{T}^{max}_{c_{\swarrow}}$, if $u_{root}.x < w(r).x_{min}$, we traverse its right child node.
Otherwise, we generate a random number $rnd_1 \in [1, \mu]$ and count the number of points that may exist in $w(r)$ from $\mathcal{B}_{root}^{max}$.
Let $cnt_{root}$ be this number.
\begin{itemize}
    \setlength{\leftskip}{-2.0mm}
    \item   If $rnd_1 \leq cnt_{root}$, we use $\mathcal{B}_{root}^{max}$.
            We pick a bucket $B \in \mathcal{B}_{root}^{max}$ with probability $\frac{\log m}{cnt_{root}}$ and then pick a random point $s \in B$.
    \item   If $rnd_1 > cnt_{root}$, we generate a random number $rnd_2 \in [1, \mu - cnt_{root}]$.
            Let $u_i$ be the right child node of $u_{root}$, and we count the number $cnt_{i}$ of points that may exist in $w(r)$ from $A_{i}^{max}$.
            \begin{itemize}
                \setlength{\leftskip}{-2.0mm}
                \item   If $rnd_2 \leq cnt_{i}$, we do random sampling in the above way.
                \item   Otherwise, we decrement $\mu$ by $cnt_{root} + cnt_{i}$ and traverse the left child node of $u_{root}$.
            \end{itemize}
\end{itemize}
We pick a sample $s$ in the above tree traversal approach.
Note that $s$ may not have $w(r) \cap s$ in case 3, so we insert $(r,s)$ into the result set iff $w(r) \cap s$.

\vs
\noindent
\textbf{Time complexity in this phase} is seen from:

\begin{lemma}   \label{lemma:sampling-time}
The sampling phase needs $O(t\log^{2}m)$ expected time.
\end{lemma}

\noindent
\textsc{Proof.}
To prove this lemma, we analyze the time to obtain one join sample $(r,s)$.
The time to pick $r$ is $O(1)$, and the time to pick a cell $c'$ is also $O(1)$.
When $c'$ falls into case 1, $(r,s)$ is obtained in $O(1)$ time.
When $c'$ falls into case 2, $(r,s)$ is obtained in $O(\log m)$ time.

When $c'$ falls into case 3, from \cref{lemma:counting-time}, we need $O(\log m)$ time to pick a point $s$.
However, we may not have $w(r) \cap s$.
Since $|J| = O(nm)$, each $r \in R$ has at least one point $s \in S$ such that $w(r) \cap s$ in average.
From \cref{lemma:bound}, with $O(\log m)$ iterations, we expect that we can pick $s$ satisfying $w(r) \cap s$ in case 3.
That is, in case 3, $(r,s)$ is obtained in $O(\log^{2}m)$ expected time.
\wsq

\subsection{Analysis}
Finally, we clarify the time complexity, space complexity, and correctness of our algorithm\footnote{Limitation remark: Although our technique can be extended for the case of higher dimensional points, this case suffers from decreased time and space efficiencies as the grid size and the complexity of BBST grow exponentially to the dimensionality.}.

\begin{theorem}[\textsc{Time complexity}]   \label{theorem:time}
\cref{algo:proposed} terminates in $O(n\log m + m\log m + t\log^{2}m)$ expected time.   
\end{theorem}

\noindent
\textsc{Proof.}
From \cref{lemma:phase1-time,lemma:counting-time,lemma:sampling-time}.
\wsq

\begin{theorem}[\textsc{Space complexity}]  \label{theorem:space}
\cref{algo:proposed} needs $O(n + m)$ space.   
\end{theorem}

\noindent
\textsc{Proof.}
We have $\sum_{c \in G}(|S(c)| + |S^{y}(c)|) = O(m)$, $\sum_{r \in R}|\mathcal{A}_r| = O(n)$, and $|\mathcal{A}| = n$.
This fact and \cref{lemma:bbst:space} prove this theorem.
\wsq

\begin{theorem}[\textsc{Correctness}]
\cref{algo:proposed} picks a spatial range join sample uniformly and independently.
\end{theorem}

\noindent
\textsc{Proof.}
To prove this theorem, we prove that a join sample is picked with probability $\frac{1}{\sum_{r' \in R}\mu(r')}$.
We sample a point $r \in R$ with probability $\frac{\mu(r)}{\sum_{r' \in R}\mu(r')}$.
Then, we pick a cell $c'$ with probability $\frac{\mu(r,c')}{\mu(r)}$.
When $c'$ falls into case 1 or 2, a point $s \in S$ is picked with probability $\frac{1}{\mu(r,c')}$, so the correctness holds in cases 1 and 2.
When $c'$ falls into case 3, a point $s \in S$ is picked at a node $u_{k}$ of a BBST.
Assume that we traversed nodes $u_{0}$, $u_1$, ..., and $u_{k}$, where $u_{0}$ is the root node.
The probability that a point $s$ is picked at $u_{k}$ is
\begin{align*}
    &\frac{\mu(r,c') - cnt_0}{\mu(r,c')} \times \frac{\mu(r,c') - cnt_0 - cnt_1}{\mu(r,c') - cnt_0} \times \cdots   \\
    &\times \frac{cnt_k}{\mu(r,c') - \sum_{i \in [0,k-1]} cnt_i} \times \frac{\log m}{cnt_k} \times \frac{1}{\log m} = \frac{1}{\mu(r,c')}.
\end{align*}
(Recall that $cnt_i$ is defined in \cref{sec:proposal:sampling}.)
Now it is clear that this theorem holds.
\wsq

\section{Experiment}    \label{sec:experiment}
This section reports our experimental results.
All experiments were conducted on a Ubuntu 22.04 LTS machine with a 2.2GHz Intel Core i9-13950HX processor and 128GB RAM.

\subsection{Setting}
\noindent
\underline{\textbf{Datasets.}}
We used the following four real-world datasets that are commonly used in spatial database works \cite{li2020lisa,liu2023efficiently,doraiswamy2020gpu,yu2021geosparkviz,shahvarani2021distributed,guo2018efficient}.
\begin{itemize}
    \setlength{\leftskip}{-2.0mm}
    \item   CaStreet \cite{castreet}:
            a set of 2,249,727 MBRs in California roads.
            We used the left-bottom point of each MBR.
    \item   Foursquare \cite{yang2020lbsn2vec++}:
            a set of 11,180,160 POIs collected from Foursquare.
    \item   IMIS \cite{imis}:
            a set of 168,240,595 spatial points collected by IMIS Hellas S.A.
    \item   NYC \cite{nyc}:
            a set of 323,598,288 taxi pick-up and drop-off GPS locations in New York City.
\end{itemize}
We normalized the coordinates of each dataset so that the domain was $[0, 10000] \times [0, 10000]$.
For each dataset, we randomly assigned each point to $R$ or $S$.
By default, $|R| \approx |S|$.

\vs
\noindent
\underline{\textbf{Evaluated algorithms.}}
Our experiments evaluated the following algorithms\footnote{Range-tree \cite{chazelle1988functional}, which needs $\tilde{O}(1)$ time for an orthogonal range counting, was also tested, but it ran out of memory before completing the index building.}.
\begin{itemize}
    \setlength{\leftskip}{-2.0mm}
    \item   \textsf{KDS}:
            The algorithm introduced in \cref{sec:baseline:kds}.
    \item   \textsf{KDS-rejection}:
            The algorithm introduced in \cref{sec:baseline:kds-r}.
    \item   \textsf{BBST}:
            Our proposed algorithm presented in \cref{sec:proposal}.
\end{itemize}
The above algorithms were single-threaded, implemented in C++, and compiled by g++ 11.3.0 with -O3 optimization.
Codes are available at \url{https://github.com/amgt-d1/RS-over-SRJ}.

Since \textsf{KDS}, \textsf{KDS-rejection}, and \textsf{BBST} guarantee uniform and independent random samples, our experiments focus on time and space efficiencies, as with existing random sampling on search and join works \cite{afshani2017independent,hu2014independent,afshani2019independent,xie2021spatial,simpler2023aoyama,tao2022algorithmic,aumuller2022sampling,zhao2020efficient}.

\vs
\noindent
\textbf{Parameters.}
To define $w(r)$, we set $w(r).x_{min} = r.x - l$, $w(r).x_{max} = r.x + l$, $w(r).y_{min} = r.y - l$, and $w(r).y_{min} = r.y + l$.
By default, we set $l = 100$ and $t = 1,000,000$.
Note that, in this setting, the join size is much smaller than $n \times m$ (only $0.0056 \times nm$ on CaStreet and $0.027 \times nm$ on IMIS for example)\footnote{We found that join sizes are still larger than at least billions.
(Recall that our problem targets large join results, and small join result size cases are not interesting for our problem.)
Thus, ``join then random sampling'' is trivially slower than the above algorithms (see \cref{sec:experiment:result}) and tends to have run out of memory.}.

\begin{table}[!t]
    \centering
    \caption{Pre-processing time [sec]}
    \label{tab:preprocessing}
    \vspace{-2.0mm}
    \begin{tabular}{lcccc} \toprule
        Dataset         & CaStreet      & Foursquare    & IMIS              & NYC               \\ \midrule
        \textsf{KDS}    & 0.36          & 1.89          & 29.09             & 48.16             \\ 
        \textsf{BBST}   & \textbf{0.14} & \textbf{0.99} & \textbf{14.23}    & \textbf{22.97}    \\ \bottomrule
    \end{tabular}
    \vspace{-2.0mm}
\end{table}

\begin{figure*}[!t]
    \begin{center}
        \subfigure[CaStreet]{%
    	\includegraphics[width=0.23\linewidth]{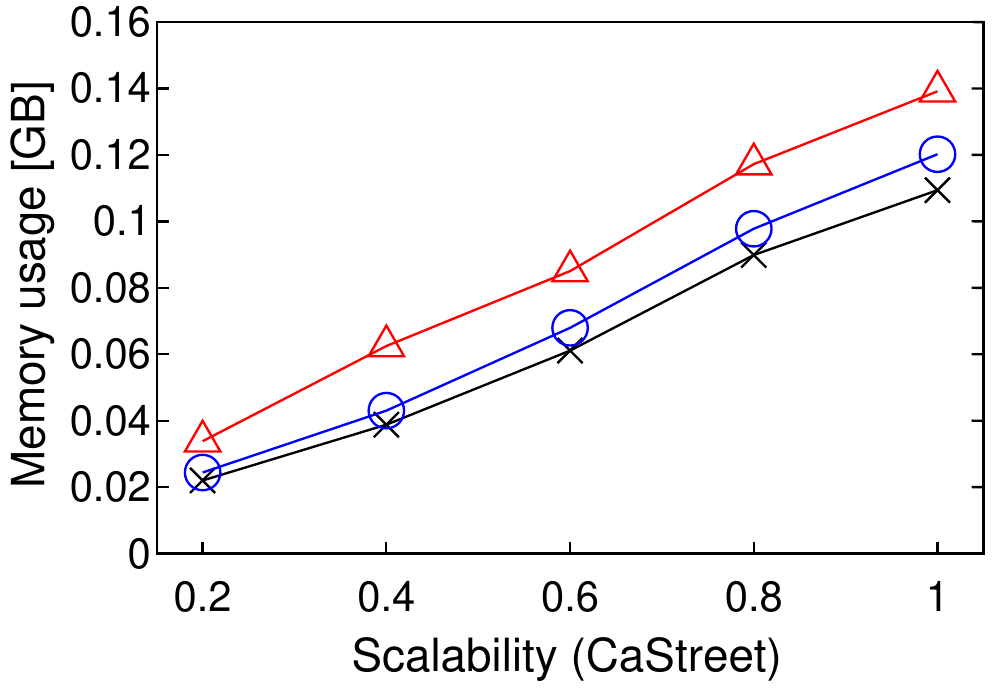}    \label{fig:castreet_memory}}
        \subfigure[Foursquare]{%
    	\includegraphics[width=0.23\linewidth]{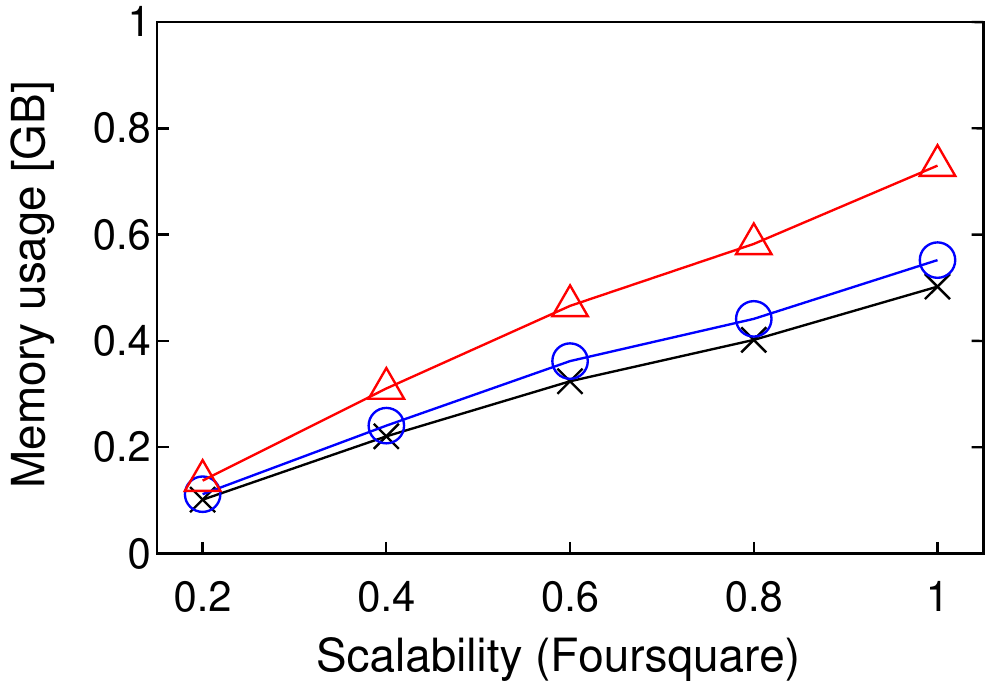}  \label{fig:foursquare_memory}}
        \subfigure[IMIS]{%
            \includegraphics[width=0.23\linewidth]{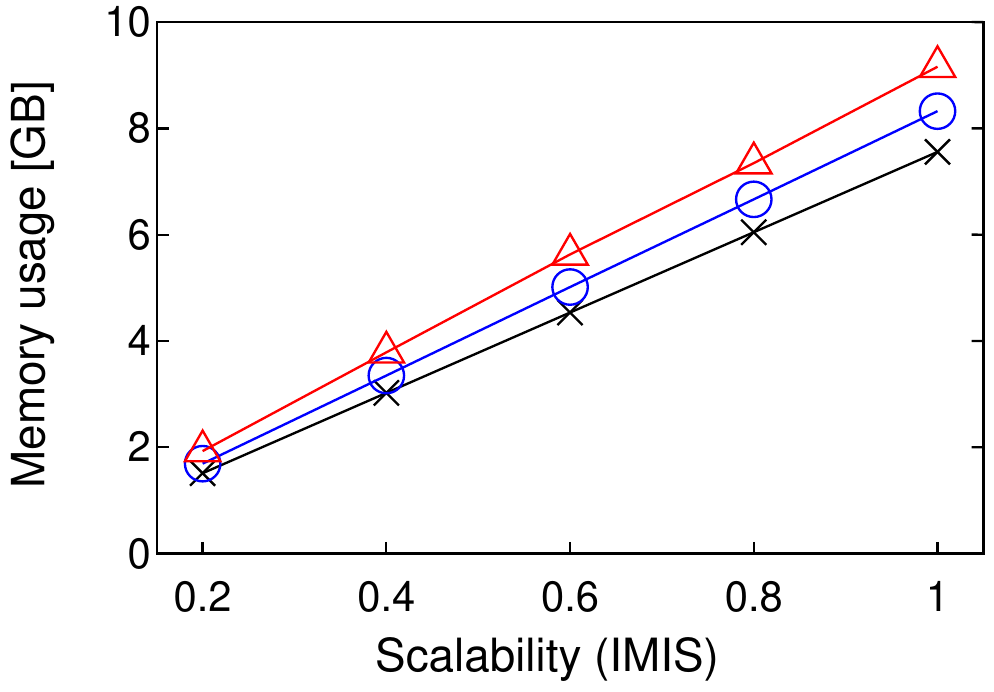}        \label{fig:imis_memory}}
        \subfigure[NYC]{%
    	\includegraphics[width=0.23\linewidth]{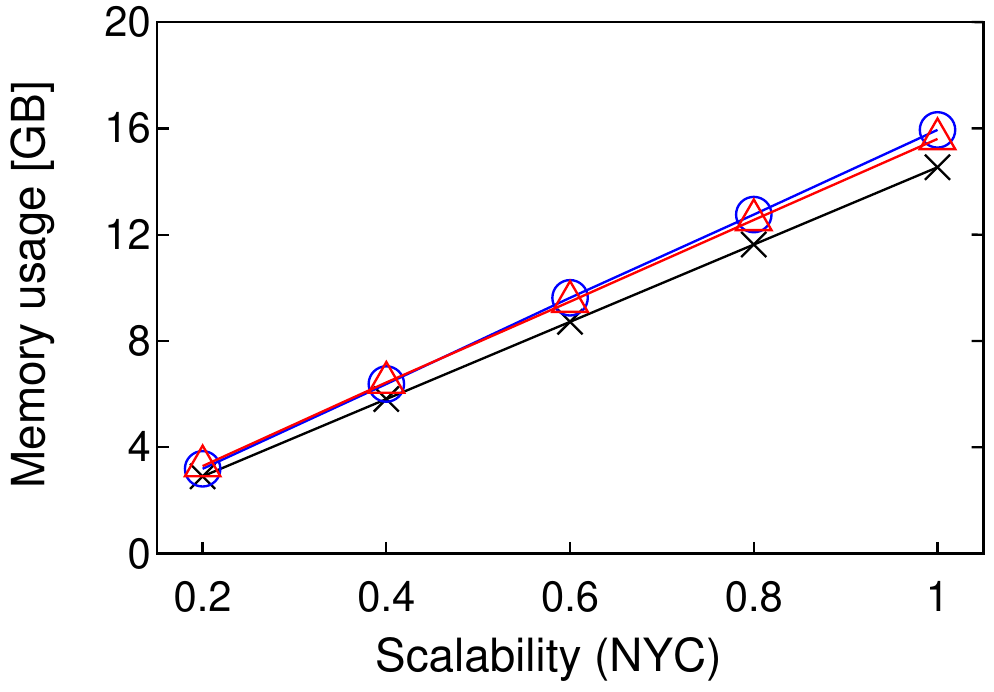}         \label{fig:nyc_memory}}
        \vspace{-2.0mm}
        \caption{Memory usage vs. dataset size.
        ``$\times$'' shows \textsf{KDS}, ``\textcolor{blue}{$\circ$}'' shows \textcolor{blue}{\textsf{KDS-rejection}}, and ``\textcolor{red}{$\triangle$}'' shows \textcolor{red}{\textsf{BBST}}.}
        \label{fig:memory}
        \vspace{-5.0mm}
    \end{center}
\end{figure*}
\begin{table*}[!t]
    \centering
    \caption{Total and decomposed times [sec]. GM and UB respectively represent grid mapping and upper-bounding. For \textsf{BBST}, GM (UB) corresponds to the online data structure building (approximate range counting) phase.}
    \label{tab:decomposed}
    \vspace{-2.0mm}
    \begin{tabular}{l|rrr|rrr|rrr|rrr} \toprule
                                & \multicolumn{3}{c|}{CaStreet}     & \multicolumn{3}{c|}{Foursquare}   & \multicolumn{3}{c|}{IMIS}             & \multicolumn{3}{c}{NYC}               \\ \hline
        Dataset                 & Total         & GM        & UB    & Total         & GM        & UB    & Total             & GM    & UB        & Total             & GM    & UB        \\ \midrule
        \textsf{KDS}            & 35.29         & -         & 3.62  & 116.90        & -         & 36.64 & 2844.62           & -     & 2261.52   & 3252.38           & -     & 555.97    \\ 
        \textsf{KDS-rejection}  & 71.10         & 0.07      & 0.10  & 132.49        & 0.32      & 0.42  & 1062.01           & 5.20  & 6.81      & 3811.34           & 8.11  & 12.48     \\ 
        \textsf{BBST}           & \textbf{1.54} & 0.20      & 0.79  & \textbf{5.67} & 1.06      & 4.11  & \textbf{72.02}    & 12.67 & 58.68     & \textbf{30.55}    & 21.74 & 7.60      \\ \bottomrule
    \end{tabular}
    \vspace{-3.0mm}
\end{table*}
\begin{table*}[!t]
    \centering
    \caption{Sampling time [sec] and \#sampling iterations ($t = 1,000,000$)}
    \label{tab:decomposed_sample}
    \vspace{-2.0mm}
    \begin{tabular}{l|rr|rr|rr|rr} \toprule
                                & \multicolumn{2}{c|}{CaStreet} & \multicolumn{2}{c|}{Foursquare}   & \multicolumn{2}{c|}{IMIS} & \multicolumn{2}{c}{NYC}   \\ \hline
        Dataset                 & Sampling  & \#iterations      & Sampling  & \#iterations          & Sampling  & \#iterations  & Sampling  & \#iterations  \\ \midrule
        \textsf{KDS}            & 31.67     & 1,000,000         & 80.26     & 1,000,000             & 583.12    & 1,000,000     & 2696.42   & 1,000,000     \\ 
        \textsf{KDS-rejection}  & 70.93     & 1,883,952         & 131.73    & 1,449,218             & 1049.99   & 1,476,050     & 3790.74   & 1,803,540     \\ 
        \textsf{BBST}           & 0.55      & 1,143,805         & 0.50      & 1,029,027             & 0.67      & 1,066,613     & 1.21      & 1,168,306     \\ \bottomrule
    \end{tabular}
    \vspace{-4.0mm}
\end{table*}

\subsection{Result} \label{sec:experiment:result}
\noindent
\underline{\textbf{Pre-processing time.}}
We first focus on the pre-processing time of each algorithm.
\cref{tab:preprocessing} shows the result.
\textsf{KD}S and \textsf{KDS-rejection} share the same pre-processing (building a $k$d-tree of $S$), so the result of \textsf{KDS-rejection} is omitted.
The pre-processing time of each algorithm is theoretically the same, i.e., $O(m\log m)$ time.
However, as \textsf{BBST} needs only sorting in pre-processing, \textsf{BBST} has a shorter pre-processing time than \textsf{KDS} (and \textsf{KDS-rejection}).

\vs
\noindent
\underline{\textbf{Memory usage.}}
\cref{fig:memory} shows the results of experiments that measured the memory usage of each algorithm by varying the dataset size (with random sampling).
As with \cref{theorem:space}, \textsf{BBST} needs a space linear to the dataset size.
Although \textsf{BBST} consumes more memory than \textsf{KDS} and \textsf{KDS-rejection} in practice, the difference is slight.
Moreover, even when $S$ has hundreds of millions of points (e.g., NYC), the memory usage is reasonable and easy to fit into modern RAMs.

Recall that we build $\mathcal{T}^{min}_{c}$ and $\mathcal{T}^{max}_{c}$ for each cell $c$ of the grid.
Clearly $|S(c)| \ll m$, so the heights of $\mathcal{T}^{min}_{c}$ and $\mathcal{T}^{max}_{c}$ are (much) less than $\log m$.
Therefore, the space costs of $A^{max}_{i}$ and $A^{min}_{i}$ are saved in practice, and this approach makes the memory usage similar to that for a single $k$d-tree.

\vs
\noindent
\underline{\textbf{Accuracy of approximate range counting.}}
We empirically clarify that our approximate range counting yields low errors.
To investigate errors, we measured $\frac{\sum_{r \in R} \mu(r)}{\sum_{r \in R}|S(w(r))|} = \frac{\sum_{r \in R} \mu(r)}{|J|}$.

In the default setting, the errors are 1.19, 1.04, 1.07, and 1.17 on CaStreet, Foursquare, IMIS, and NYC, respectively.
This result confirms that our upper-bounding empirically yields much better accuracy than the $O(\log m)$ bound.
Thanks to this effectiveness, the number of sampling iterations in the sampling phase is nearly $t$, which will be seen later.

\vs
\noindent
\underline{\textbf{Comparison with baselines.}}
We compared \textsf{BBST} with \textsf{KDS} and \textsf{KDS-rejection}.
\cref{tab:decomposed,tab:decomposed_sample} detail the computation times of these algorithms at the default setting.
As a first observation, the total time result showcases that \textsf{BBST} is at least 10 times faster than \textsf{KDS} and \textsf{KDS-rejection}.

To further understand the performance difference between \textsf{BBST} and the baselines, we compare \textsf{BBST} with \textsf{KDS} based on GM (Grid Mapping), UB (Upper-Bounding), and sampling times.
Since \textsf{KDS} does not use the grid structure, its GM is blank.
Note that GM and UB of \textsf{BBST} respectively correspond to the online data structure building and approximate range counting phases.
W.r.t. UB, \textsf{KDS} computes the exact range counts.
This approach incurs a much longer computation time than our approximate range counting, although the error of our upper-bounding is small.
As for sampling, \textsf{KDS} always picks $(r,s)$ such that $w(r) \cap s$, so its \#iterations is $t$.
However, \textsf{KDS} needs $O(\sqrt{m})$ time for one-time sampling, and its sampling time is not of interactive level.
On the other hand, the sampling time of \textsf{BBST} is only about 1 second, which is significantly faster than that of \textsf{KDS}.
This result is derived from two facts.
One is the high accuracy of upper-bounding, as seen before.
This high accuracy minimizes the number of failure iterations.
It can be seen that the number of iterations of \textsf{BBST} is $\approx t \times \frac{\sum_{r \in R} \mu(r)}{|J|}$.
Furthermore, assume that $|S(w(r))|$ is pre-known for each $r \in R$, that is, \textsf{KDS} only does sampling.
Even in this known selectivity case, \textsf{BBST} is still faster than \textsf{KDS}, see the total time of \textsf{BBST} and the samplings time of \textsf{KDS}.
(This also holds for \textsf{KDS-rejection}.)

\begin{figure*}[!t]
    \begin{center}
        \subfigure[CaStreet]{%
    	\includegraphics[width=0.23\linewidth]{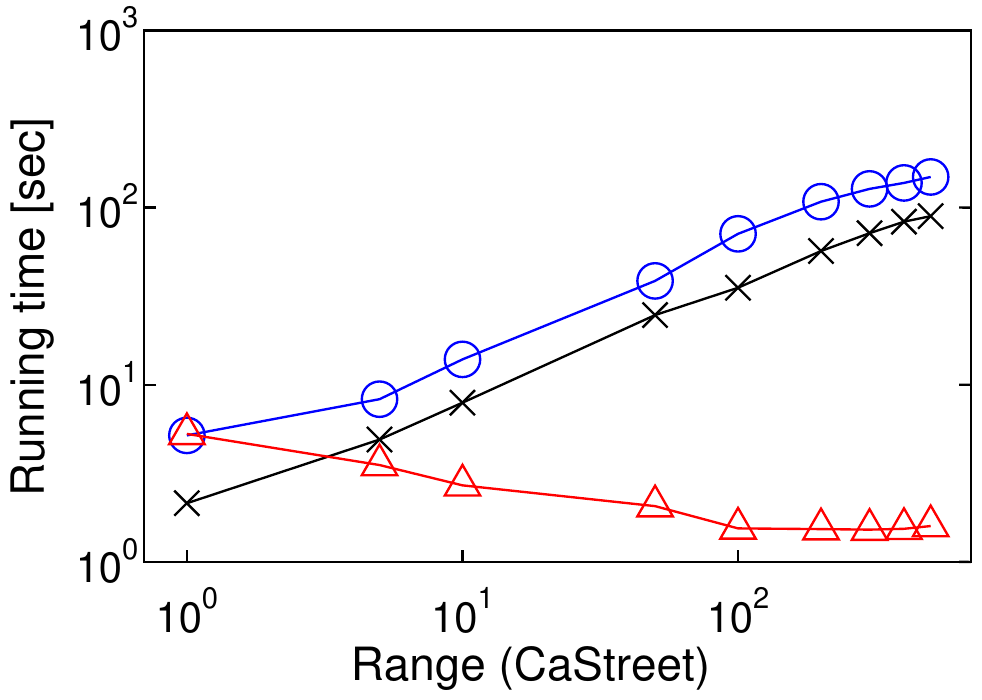}     \label{fig:castreet_range}}
        \subfigure[Foursquare]{%
    	\includegraphics[width=0.23\linewidth]{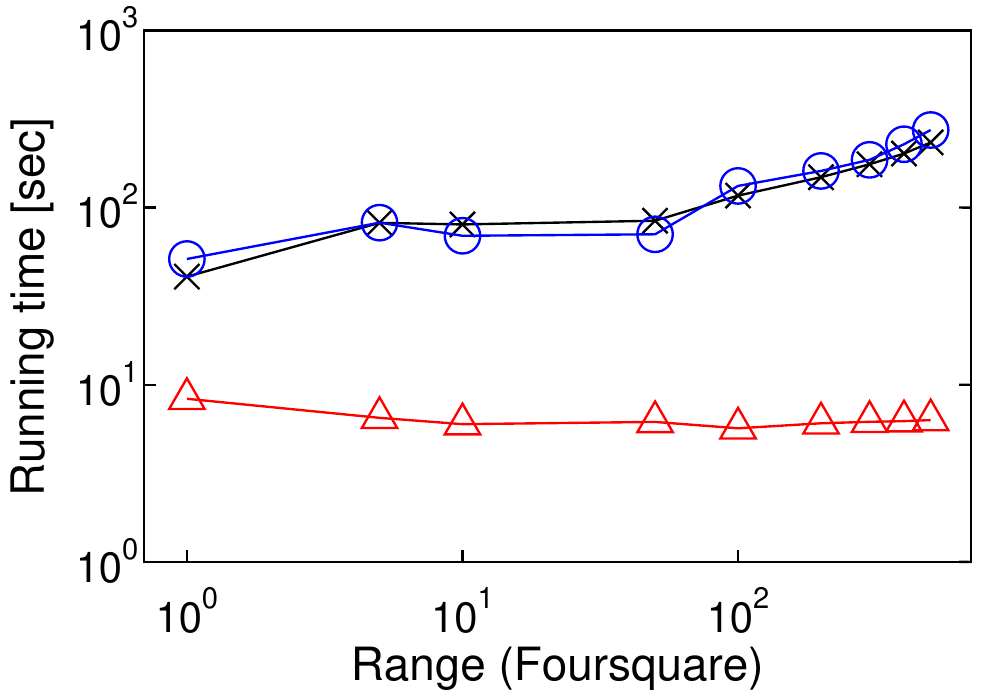}   \label{fig:foursquare_range}}
        \subfigure[IMIS]{%
            \includegraphics[width=0.23\linewidth]{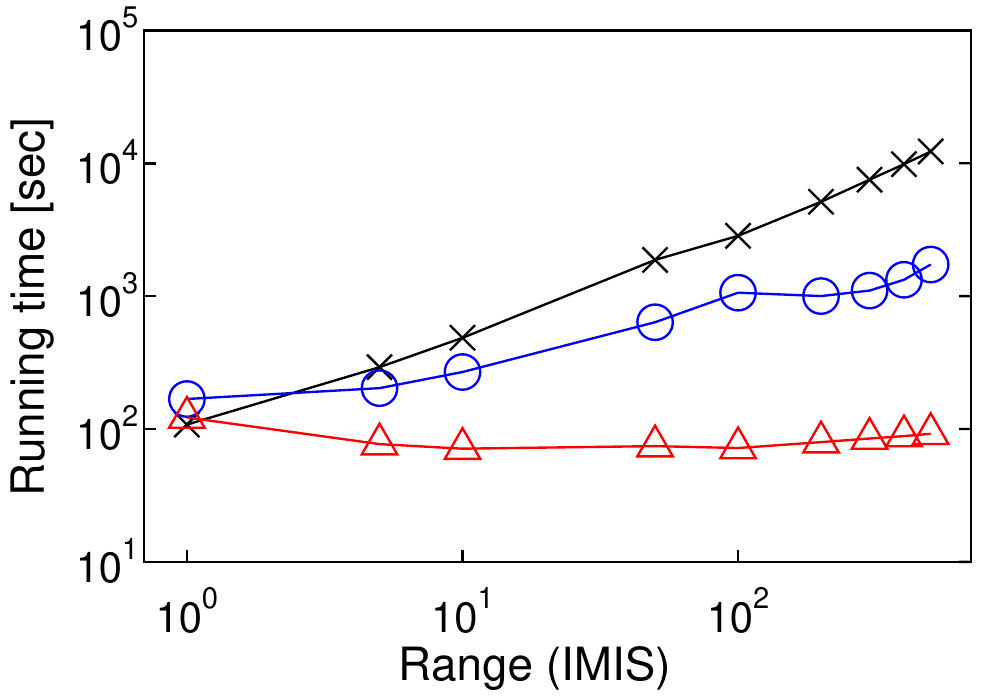}         \label{fig:imis_range}}
        \subfigure[NYC]{%
    	\includegraphics[width=0.23\linewidth]{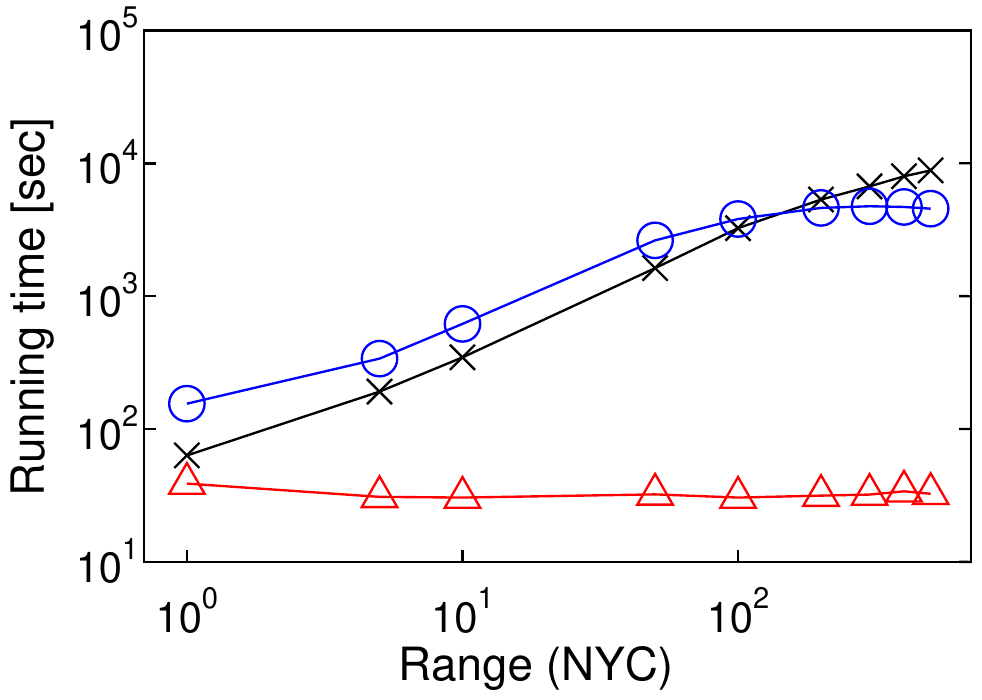}          \label{fig:nyc_range}}
        \vspace{-2.0mm}
        \caption{Impact of range (window) size.
        ``$\times$'' shows \textsf{KDS}, ``\textcolor{blue}{$\circ$}'' shows \textcolor{blue}{\textsf{KDS-rejection}}, and ``\textcolor{red}{$\triangle$}'' shows \textcolor{red}{\textsf{BBST}}.}
        \label{fig:range}
        \vspace{-6.0mm}
    \end{center}
\end{figure*}
\begin{figure*}[!t]
    \begin{center}
        \subfigure[CaStreet]{%
    	\includegraphics[width=0.23\linewidth]{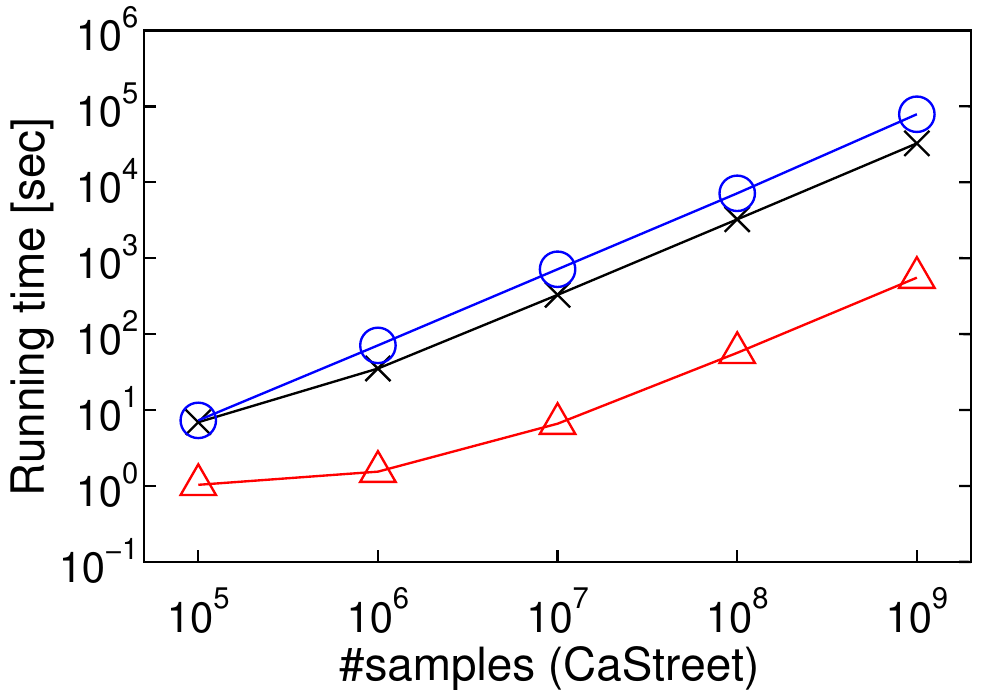}    \label{fig:castreet_sample}}
        \subfigure[Foursquare]{%
    	\includegraphics[width=0.23\linewidth]{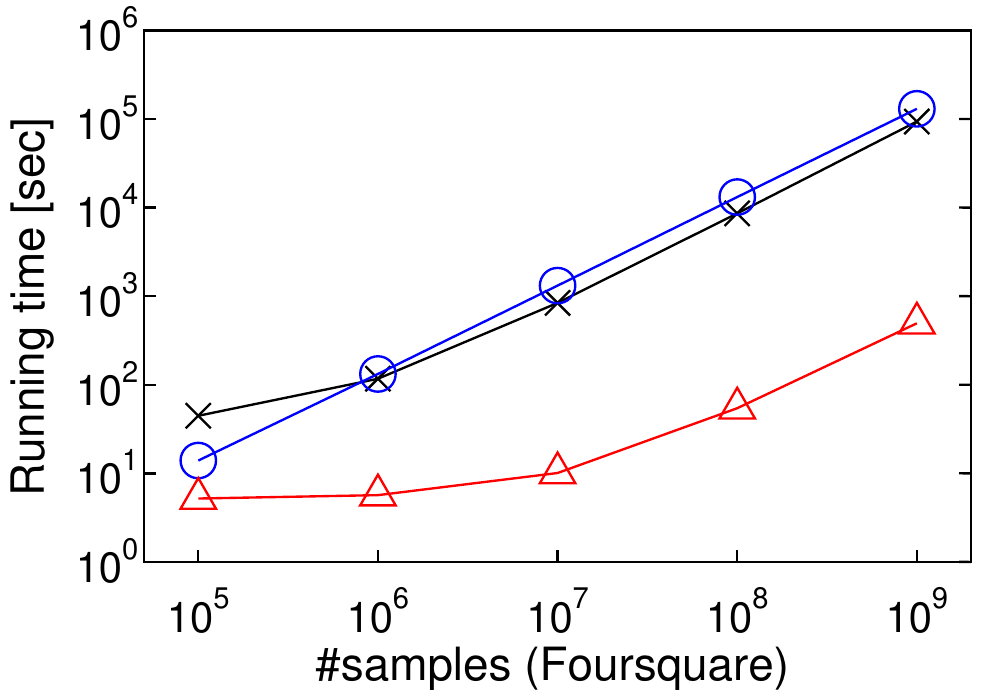}  \label{fig:foursquare_sample}}
        \subfigure[IMIS]{%
            \includegraphics[width=0.23\linewidth]{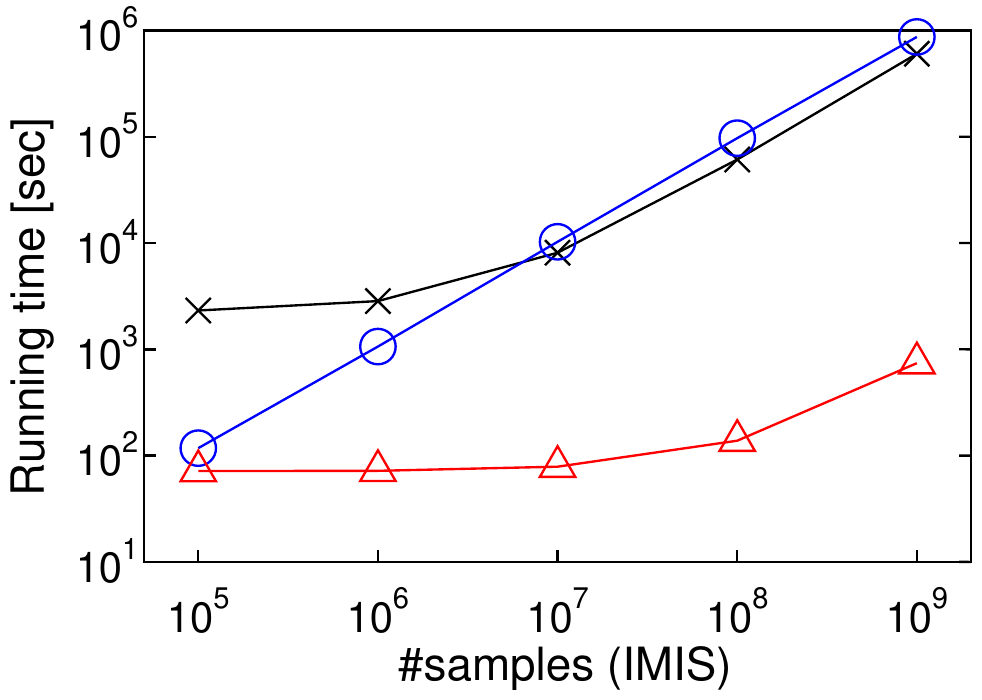}        \label{fig:imis_sample}}
        \subfigure[NYC]{%
    	\includegraphics[width=0.23\linewidth]{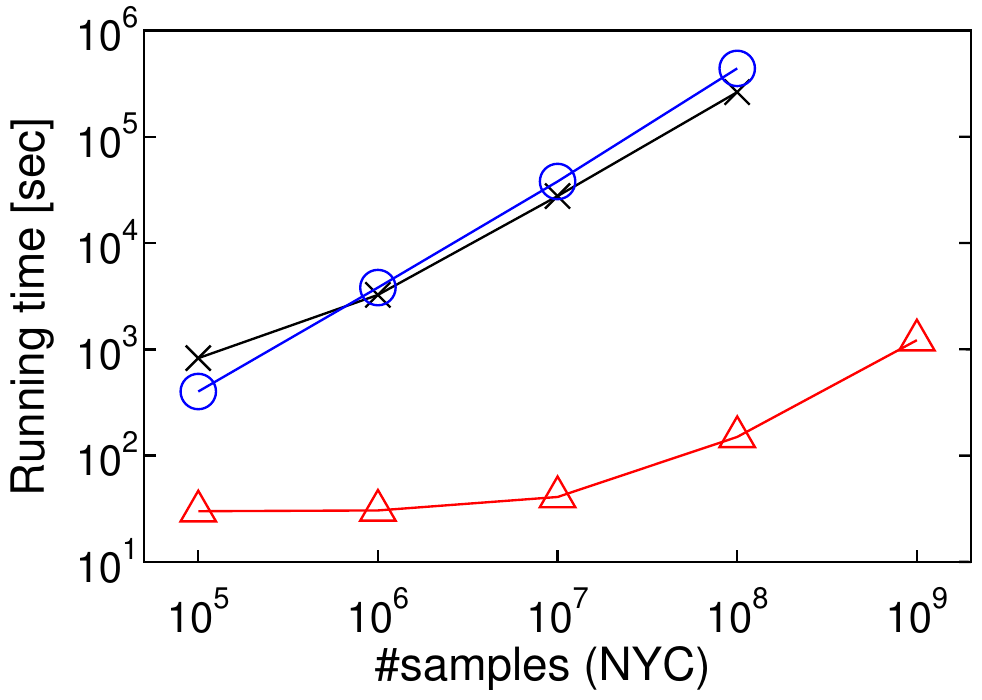}         \label{fig:nyc_sample}}
        \vspace{-2.0mm}
        \caption{Impact of \#samples.
        ``$\times$'' shows \textsf{KDS}, ``\textcolor{blue}{$\circ$}'' shows \textcolor{blue}{\textsf{KDS-rejection}}, and ``\textcolor{red}{$\triangle$}'' shows \textcolor{red}{\textsf{BBST}}.}
        \label{fig:sample}
        \vspace{-5.0mm}
    \end{center}
\end{figure*}

Next, we compare \textsf{BBST} with \textsf{KDS-rejection}.
We see that the GM time of \textsf{BBST} is longer than that of \textsf{KDS-rejection}.
This is because \textsf{BBST} builds $\mathcal{T}^{min}_{c}$ and $\mathcal{T}^{max}_{c}$ for each cell $c \in G$ in addition to mapping each point into its corresponding cell.
Furthermore, the UB time of \textsf{KDS-rejection} usually is shorter than that of \textsf{BBST}, because \textsf{KDS-rejection} needs only $O(n)$ time for upper-bounding.
However, because the upper-bounding approach of \textsf{KDS-rejection} yields loose upper-bounds, \textsf{KDS-rejection} suffers from a low acceptance probability in its rejection sampling.
\textsf{KDS-rejection} consequently faces the inefficiency of the sampling algorithm and a large number of sampling iterations.
This result clarifies that, although both \textsf{BBST} and \textsf{KDS-rejection} employ a grid structure, our approach better exploits this structure.

Furthermore, notice that UB time of \textsf{KDS} is longer than the total time of \textsf{BBST}.
As UB of \textsf{KDS} is to \textit{count} the join size, it is obvious that running a spatial join incurs longer time than the UB time of \textsf{KDS}.
Therefore, we see that \textsf{BBST} is faster than spatial join algorithms.

\vs
\noindent
\underline{\textbf{Impact of range (window) size.}}
To investigate the influence of $|J|$, we varied $l$ (the range or window size) from 1 to 500.
A larger $l$ yields a larger $|J|$.
\cref{fig:range} depicts the experimental results.
We observe that \textsf{BBST} can be worse than \textsf{KDS} when $l$ is significantly small (e.g, $l = 1$) although the difference is small.
When the window size is small, many cells are created, necessitating longer GM and UB times.
However, \textsf{BBST} is not very sensitive to the window size.
This result is theoretically reasonable because the search spaces of our approaches for the approximate range counting and sampling phases are not affected by the range size.
\cref{lemma:counting-time,lemma:sampling-time} theoretically support this observation (although \textsf{BBST} is affected by the constant factor owing to a given data distribution).
This is one of the merits of \textsf{BBST}, because the performance of \textsf{BBST} does not degrade even when $|J|$ is huge.

The running times of \textsf{KDS} and \textsf{KDS-rejection} are affected by the range size, which is consistent with the result in \cite{xie2021spatial}.
We found that the times for range counting and sampling by a $k$d-tree increase as $l$ increases.
Again, our new data structure BBST does not have this drawback.

\begin{figure*}[!t]
    \begin{center}
        \subfigure[CaStreet]{%
    	\includegraphics[width=0.23\linewidth]{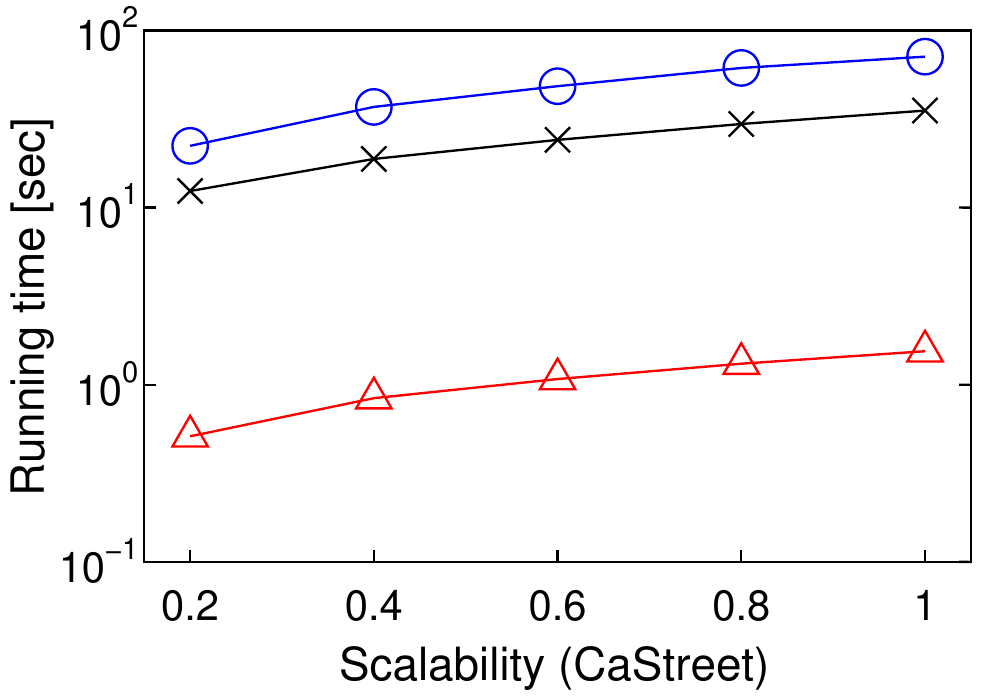}   \label{fig:castreet_scalability}}
        \subfigure[Foursquare]{%
    	\includegraphics[width=0.23\linewidth]{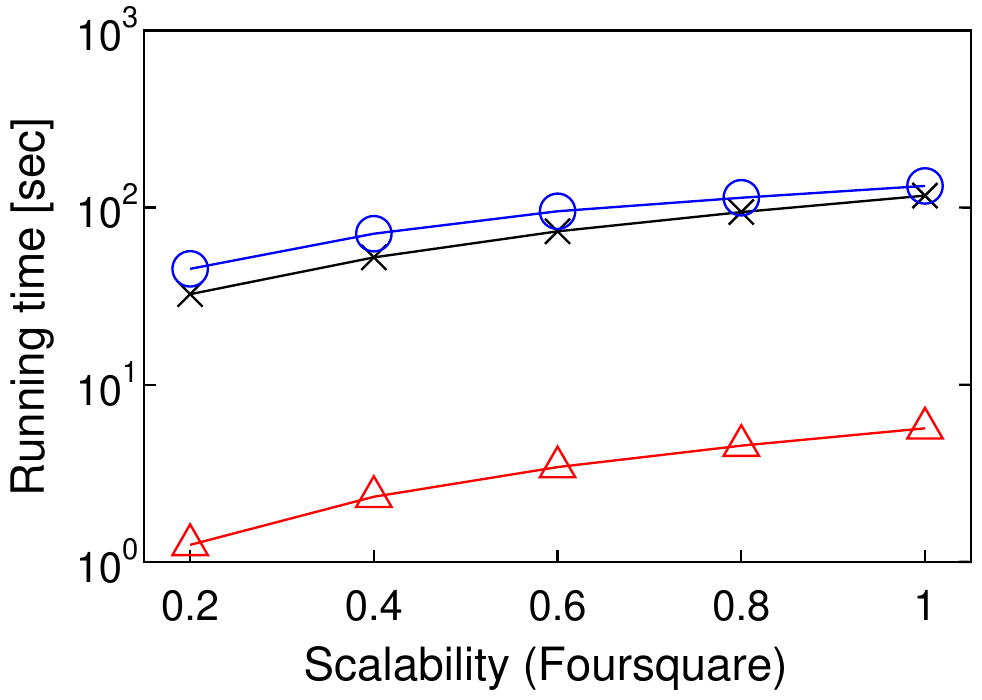} \label{fig:foursquare_scalability}}
        \subfigure[IMIS]{%
            \includegraphics[width=0.23\linewidth]{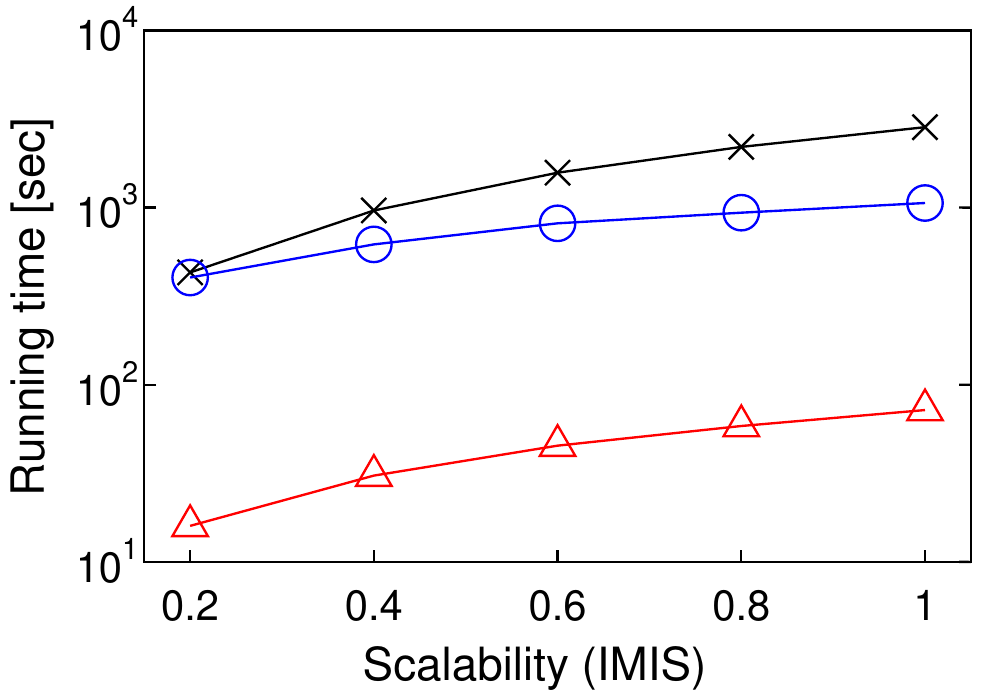}       \label{fig:imis_scalability}}
        \subfigure[NYC]{%
    	\includegraphics[width=0.23\linewidth]{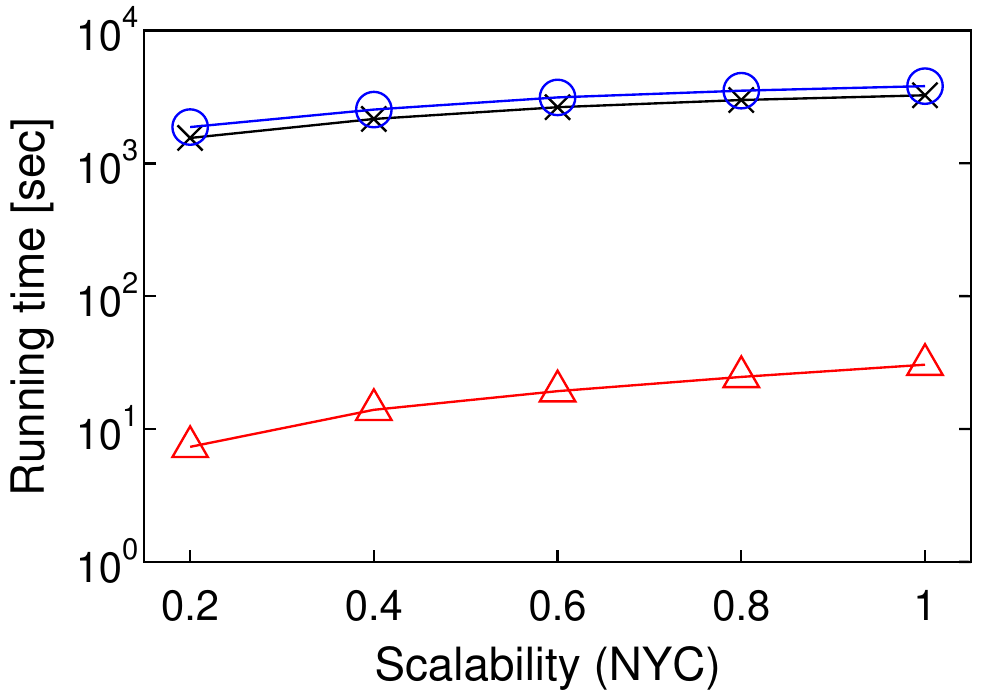}        \label{fig:nyc_scalability}}
        \vspace{-2.0mm}
        \caption{Impact of dataset size.
        ``$\times$'' shows \textsf{KDS}, ``\textcolor{blue}{$\circ$}'' shows \textcolor{blue}{\textsf{KDS-rejection}}, and ``\textcolor{red}{$\triangle$}'' shows \textcolor{red}{\textsf{BBST}}.}
        \label{fig:cardinality}
        \vspace{-3.0mm}
    \end{center}
\end{figure*}

\vs
\noindent
\underline{\textbf{Impact of \#samples.}}
Next, we investigate how $t$ affects the practical performance of each algorithm.
\cref{fig:sample} exhibits the result.
Recall \cref{tab:decomposed,tab:decomposed_sample}.
The dominant costs of \textsf{KDS} and \textsf{KDS-rejection} are basically derived from sampling.
Hence, their running times increase linearly to $t$.

On the other hand, the running time of \textsf{BBST} does not increase linearly to $t$ when $t \leq 1,000,000,000$.
Again, recall \cref{tab:decomposed,tab:decomposed_sample}.
The cost of sampling can be dominant only when $t$ is sufficiently large.
Therefore, the running time of \textsf{BBST} \textit{gradually} increases with the increase of $t$.
This is also one of the merits of \textsf{BBST}: the performance difference between \textsf{BBST} and the baselines becomes larger as $t$ increases.
For example, \textsf{BBST} picked 100 million join samples within about a few minutes on NYC, whereas \textsf{KDS} and \textsf{KDS-rejection} needed more than 72 hours.
(We terminated the experiments of \textsf{KDS} and \textsf{KDS-rejection} on NYC with $t = 1,000,000,000$, because they did not finish within two weeks.)

\vs
\noindent
\underline{\textbf{Impact of dataset size.}}
We investigated the scalability of each algorithm to the dataset size with the same setting for the memory usage experiment.
\cref{fig:cardinality} shows the experimental result.
\textsf{BBST} always outperforms \textsf{KDS} and \textsf{KDS-rejection}.
Now it is clear that \textit{\textsf{BBST} is superior to these baselines w.r.t. all input parameters in \cref{definition:problem}}.

\begin{figure}[!t]
    \begin{minipage}[t]{0.485\linewidth}
        \centering
        \includegraphics[width=1.0\linewidth]{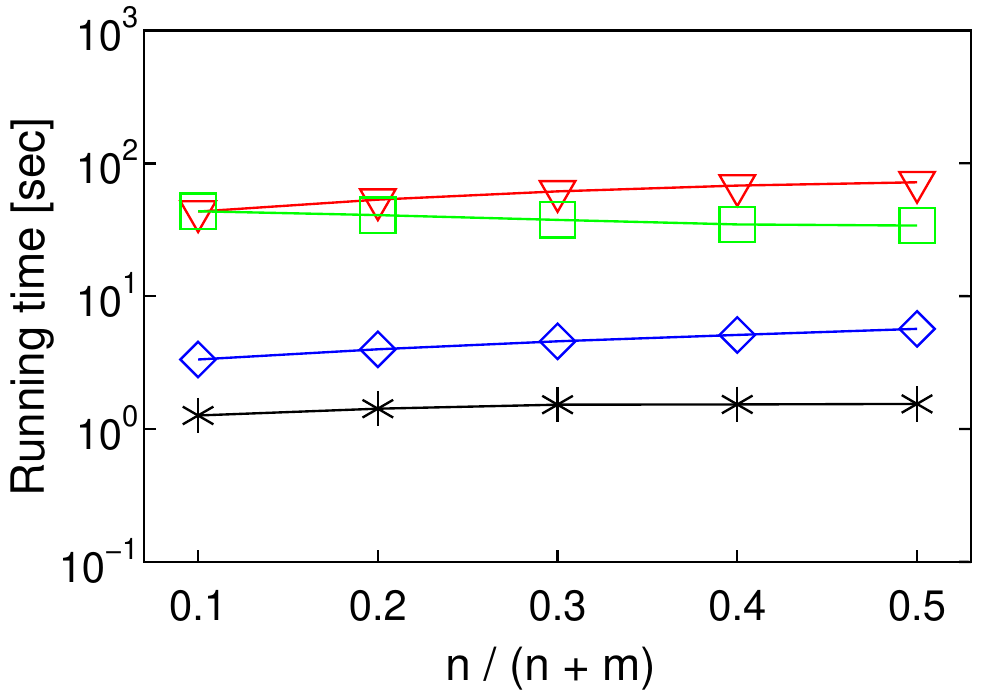}
        \vspace{-5.0mm}
        \caption{Impact of dataset size difference.
        ``$\ast$'', ``\textcolor{blue}{$\diamond$}'', ``\textcolor{red}{$\triangledown$}'', and ``\textcolor{green}{$\square$}'' show the results on CaStreet, \textcolor{blue}{Foursquare}, \textcolor{red}{IMIS}, and \textcolor{green}{NYC}, respectively.}
        \label{fig:ratio}
    \end{minipage}
    \hspace{1.0mm}
    \begin{minipage}[t]{0.485\linewidth}
        \centering
        \includegraphics[width=1.0\linewidth]{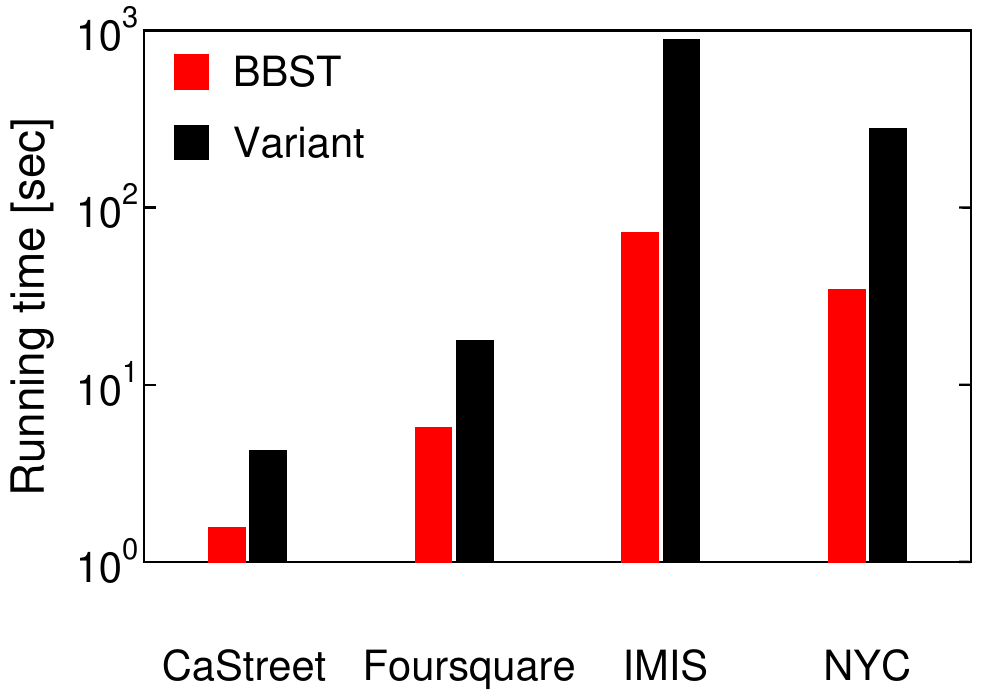}
        \caption{Comparison between BBST and $k$d-tree structures}
        \label{fig:variant}
    \end{minipage}
    \vspace{-3.0mm}
\end{figure}

\vs
\noindent
\underline{\textbf{Impact of dataset size difference.}}
We studied how the size ratio between $R$ and $S$ affects the performance of \textsf{BBST}.
(From the other experimental results, it is clear that \textsf{BBST} outperforms \textsf{KDS} and \textsf{KDS-rejection}, so we show only the results of \textsf{BBST}.)
\cref{fig:ratio} depicts the result.
Recall that $R$ and $S$ are symmetric, so we varied $n / (n+m) = |R| / (|R| + |S|)$ from 0.1 to 0.5 (as $n$ increases, $m$ decreases).

The result shows that the running time of \textsf{BBST} is almost stable or increases as $n$ increases.
\cref{lemma:counting-time} clarifies that as $n$ becomes larger, the cost of UB becomes larger.
Also, \cref{lemma:bbst:build-time} suggests that as $m$ becomes larger (i.e., $n$ becomes fewer in the experiment), the cost of GM becomes larger.
As seen in \cref{tab:decomposed}, on Castreet, Foursquare, and IMIS, UB is the main bottleneck, so the running time of \textsf{BBST} increases a bit on these datasets as $n$ increases.
On NYC, GB is the main bottleneck, and the running time does not increase.

\vs
\noindent
\underline{\textbf{Effectiveness of BBST structure.}}
To show the effectiveness of the BBST structure for our problem, we prepared a variant of \cref{algo:proposed} that employs a $k$d-tree instead of two BBSTs for each cell.
For random sampling in case 3, this variant used KDS.
\cref{fig:variant} depicts the comparison result.
We found that \textsf{BBST} outperforms the variant and is up to 12 times faster, confirming its superiority w.r.t. spatial range join sampling.

\section{Related Work}  \label{sec:related-work}
\noindent
\underline{\textbf{Spatial join.}}
The spatial range join problem is important in many applications, so many works devised efficient algorithms for this problem.
The index nested-loop join algorithm \cite{jacox2007spatial} is a simple yet still state-of-the-art approach \cite{gu2023rlr}.
Another state-of-the-art approach is based on plane-sweep \cite{patel1996partition}.
These approaches were experimentally compared in \cite{sowell2013experimental,vsidlauskas2014spatial,nobari2017memory}.
A machine learning model for optimizing spatial joins was proposed in \cite{vu2021learned,vu2024learning}.
Notice that a spatial join algorithm incurs $\Omega(|J|)$ time \cite{wang2024optimal}, and $|J|$ can be as large as $nm$.
Therefore, spatial range join algorithms cannot solve our problem efficiently.

Spatial joins have been considered in other settings.
For example, literature \cite{sabek2017spatial} considers spatial joins in MapReduce environments, whereas literature \cite{whitman2019distributed} assumes a Spark environment.
Such distributed techniques were also compared in \cite{pandey2018good}.
Streaming joins were studied in \cite{shahvarani2021distributed}, and algorithms for intersection joins on polygon data were developed in \cite{georgiadis2023raster,zacharatou2017gpu}.
These settings are totally different from our problem setting.

\vs
\noindent
\underline{\textbf{Range counting}} also has many existing studies, e.g., \cite{nekrich2014efficient,rahul2017approximate,chan2016adaptive,shekelyan2021approximating}.
Unfortunately, these existing data structures are specific to spatial range counting, and they do not have a function that can access a random point, which is involved in the range counting, uniformly at random.
Our BBST structure does not have this limitation and has a strong theoretical guarantee for random sampling over spatial joins.

\vs
\noindent
\underline{\textbf{Join sampling.}}
Chaudhuri et al. \cite{chaudhuri1999random} and Acharya et al. \cite{acharya1999join} introduced join sampling.
Due to its importance, this topic has been receiving considerable attention recently.
Note that this topic has been studied in relational database settings, and existing works assume equi-joins.
After this concept was proposed, many works (e.g., \cite{haas1999ripple,li2016wander,li2017wander,li2019wander,shanghooshabad2021pgmjoins,zhao2020efficient}) proposed empirically-efficient algorithms for join sampling (with and without uniform and independent requirements).
A theoretical progress appeared in 2018, and Zhao et al. \cite{zhao2018random} reduced the worst-case time of \cite{chaudhuri1999random}.
The next progress was reported in \cite{chen2020random,carmeli2020answering,carmeli2022answering}, and the current state-of-the-art algorithm was proposed in \cite{wang2024join}.

\vs
\noindent
\underline{\textbf{Independent range sampling.}}
Given a range and the number of samples $t$, an IRS (independent range sampling) query retrieves $t$ samples from a set of points falling into the range.
This topic also has been studied by many works with diverse settings, e.g., 1-dimensional points \cite{hu2014independent,afshani2017independent}, low-dimensional points \cite{afshani2019independent,xie2021spatial,amagata2024independent,amagata2024independent_}, and high-dimensional points \cite{har2019near,aumuller2020fair,aumuller2021fair,aumuller2022sampling_,aumuller2022sampling,simpler2023aoyama}.
A nice summary of state-of-the-art IRS techniques appears in \cite{tao2022algorithmic}.
Our algorithm and the baseline ones employ this concept when obtaining join samples.

KDS \cite{xie2021spatial} is the work most related to ours since \cite{xie2021spatial} considers orthogonal ranges.
\cref{sec:baseline} showed how to employ KDS in the problem of join sampling over spatial range joins.
In \cref{sec:experiment}, we demonstrated that our algorithm always outperforms the KDS-based baseline algorithms.

\section{Conclusion}    \label{sec:conclusion}
The spatial range join problem has a number of applications and is still being studied.
Because recent spatial databases contain many points, a spatial range join returns a substantial-sized result.
This fact poses two issues: the computational time is too long and subsequent applications are overwhelmed by the result size.
To overcome these issues, this paper introduced the problem of random sampling over spatial range joins.
We proposed a new algorithm for this problem and proved that its time and space complexities are respectively $\tilde{O}(n+m+t)$ and $O(n+m)$, where $n$ and $m$ are the sizes of the input datasets.
Furthermore, we conducted extensive experiments by using four real datasets to study the practical performance of our algorithm.
The experimental results confirm that our algorithm outperforms the baselines by a large margin in most tests.

\section*{Acknowledgements}
This work was partially supported by JSPS KAKENHI Grant Number 24K14961, AIP Acceleration Research JPMJCR23U2, and Adopting Sustainable Partnerships for Innovative Research Ecosystem JPMJAP2328, JST.

\bibliographystyle{IEEEtran}
\bibliography{sample}

\end{document}